\documentclass[11pt]{article}
\pdfoutput=1

\usepackage{jcappub}

\usepackage{amsthm,graphicx,multirow}
\usepackage{epsfig}
\usepackage{latexsym, amssymb} 
\usepackage{amsmath}
\usepackage{epstopdf}
\usepackage[perpage]{footmisc}
\usepackage{soul}
\usepackage{hyperref}
\usepackage[dvipsnames]{xcolor}
\newcommand{\dd}{\mathrm{d}}

\begin{document}

\title{Scalar-tensor theories of gravity, neutrino physics, and the $H_0$ tension}

\author[a,b,c,d,1]{Mario Ballardini,}
\author[a,b,c]{Matteo Braglia,}
\author[b,c]{Fabio Finelli,}
\author[b,c]{Daniela Paoletti,}
\author[e,f]{Alexei A. Starobinsky,}
\author[g]{Caterina Umilt\`a\note{Corresponding author.}}

\affiliation[a]{Dipartimento di Fisica e Astronomia, Alma Mater Studiorum Universit\`a di Bologna, via Gobetti 93/2, I-40129 Bologna, Italy}
\affiliation[b]{INAF/OAS Bologna, via Piero Gobetti 101, I-40129 Bologna, Italy}
\affiliation[c]{INFN, Sezione di Bologna, via Irnerio 46, I-40126 Bologna, Italy}
\affiliation[d]{Department of Physics \& Astronomy, University of the Western Cape, Cape Town 7535, South Africa}
\affiliation[e]{Landau Institute for Theoretical Physics, 119334 Moscow, Russia}
\affiliation[f]{Bogolyubov Laboratory of Theoretical Physics, Joint Institute for Nuclear Research, 141980 Dubna, Moscow Region, Russia}
\affiliation[g]{Department of Physics, University of Illinois Urbana-Champaign, Urbana, IL 61801}

\emailAdd{mario.ballardini@inaf.it}
\emailAdd{matteo.braglia2@unibo.it}
\emailAdd{fabio.finelli@inaf.it}
\emailAdd{daniela.paoletti@inaf.it}
\emailAdd{alstar@landau.ac.ru}
\emailAdd{umilta@illinois.edu}

\abstract{
We use {\em Planck} 2018 data to constrain the simplest models of scalar-tensor theories 
characterized by a coupling to the Ricci scalar of the type $F(\sigma) R$ with 
$F(\sigma) = N_{pl}^2 + \xi \sigma^2$.
We update our results with previous {\em Planck} and BAO data releases obtaining the tightest 
constraints to date on the coupling parameters, that is $\xi < 5.5 \times 10^{-4}$ for $N_{pl}=0$ 
(induced gravity or equivalently extended Jordan-Brans-Dicke) and 
$(N_{pl} \sqrt{8 \pi G})-1 < 1.8 \times 10^{-5}$ for $\xi = -1/6$ (conformal coupling), both at 95\% CL.
Because of a modified expansion history after radiation-matter equality 
compared to the $\Lambda$CDM model, all these dynamical models accommodate a higher value 
for $H_0$ and therefore alleviate the tension between {\em Planck}/BAO and 
distance-ladder measurement from SNe Ia data from $4.4\sigma$ at best to 
2.7-3.2$\sigma$ with CMB alone and 3.5-3.6$\sigma$ including BAO data.
We show that all these results are robust to changes in the neutrino physics. 
In comparison to the $\Lambda$CDM model, partial degeneracies between neutrino physics and the coupling 
to the Ricci scalar allow for smaller values 
$N_{\rm eff} \sim 2.8$, $1\sigma$ lower compared to the 
standard $N_{\rm eff} = 3.046$, and relax the upper limit on the neutrino mass up to 40\%.}

    \maketitle

\section{Introduction}

The distance-ladder measurement of the Hubble constant from supernovae type Ia (SNe Ia) 
\cite{Riess:2016jrr,Riess:2018byc} and from strong-lensing time delays 
\cite{Bonvin:2016crt,Birrer:2018vtm,Wong:2019kwg} 
disagrees with the value inferred from the fit of the cosmic microwaves background (CMB) 
anisotropies \cite{Ade:2015xua,Aghanim:2018eyx}. 
In particular, the value inferred using the {\em Planck} legacy data for a flat $\Lambda$CDM 
cosmological model, $H_0 = (67.36 \pm 0.54)$ km s$^{-1}$Mpc$^{-1}$ \cite{Aghanim:2018eyx}, 
is in a 4.4$\sigma$ tension with the most recent measurement from the SH0ES team \cite{Riess:2019cxk}, 
that is $H_0 = (74.03 \pm 1.42)$ km s$^{-1}$Mpc$^{-1}$, determined using Cepheid-calibrated 
SNe Ia with new parallax measurements from HST spatial scanning \cite{Riess:2018byc} and 
from Gaia DR2 \cite{Riess:2018uxu}, in a 3.2$\sigma$ tension with the strong-lensing 
time delay determination from the H0LiCOW collaboration \cite{Wong:2019kwg}, that is 
$H_0 = \left(73.3_{-1.8}^{+1.7}\right)$ km s$^{-1}$Mpc$^{-1}$, and in a 5.3$\sigma$ tension 
with the combined SH0ES + H0LiCOW value $H_0 = (73.8 \pm 1.1)$ km s$^{-1}$Mpc$^{-1}$ \cite{Wong:2019kwg}.

Although it does not seem possible to completely solve the discrepancy between CMB and 
local measurements by considering unaccounted systematic effects 
\cite{Addison:2015wyg,Aghanim:2016sns,Aylor:2018drw}, revisions of the determination of 
the Hubble rate based on the Cepheid calibration 
\cite{Efstathiou:2013via,Rigault:2014kaa,Rigault:2018ffm}, and from SNe Ia calibrated using 
the tip of the red giant branch method \cite{Freedman:2019jwv} point to values which are 
still higher than the {\em Planck} measurements, i.e. $\sim 70$ km s$^{-1}$Mpc$^{-1}$, but to a 
smaller extent (see however \cite{Jones:2018vbn,Rose:2019ncv,Yuan:2019npk,Dhawan:2020xmp,Freedman:2020dne} 
for more recent developments). 

Alternatively, the $H_0$ {\em tension} can be addressed by evoking new physics beyond the 
$\Lambda$CDM concordance model, either in the late or early time Universe 
\cite{Mortsell:2018mfj,Knox:2019rjx}. 
Late-time modifications include a phantom-like dark energy (DE) component 
\cite{DiValentino:2016hlg,DiValentino:2017zyq,Vagnozzi:2019ezj}, a vacuum phase transition 
\cite{DiValentino:2017rcr,Khosravi:2017hfi,Banihashemi:2018has,Banihashemi:2018oxo,Benevento:2020fev}, 
interacting dark energy \cite{Kumar:2016zpg,DiValentino:2017iww,Yang:2018euj,DiValentino:2019jae,Gomez-Valent:2020mqn}, 
or quintessence in a non-flat Universe \cite{Miao:2018zpw}.
However, these models are tightly constrained 
\cite{Riess:2016jrr,DiValentino:2017zyq,Addison:2017fdm,DiValentino:2017iww} 
by late-time observational data, especially those from baryon acoustic oscillations (BAO) 
\cite{Beutler:2011hx,Ross:2014qpa,Alam:2016hwk}.  

With the highly precisely determined angular scale of the last-scattering surface 
(LSS) \cite{Akrami:2018vks}, it has been suggested that a smaller value of the comoving sound 
horizon at baryon drag $r_s$ can provide a higher value of $H_0$ without spoiling the CMB angular 
power spectrum measurements and without changing the BAO observables \cite{Bernal:2016gxb,Knox:2019rjx}.
An example of such an early-time modification  is the extension to additional light relics, 
eventually interacting with hidden dark sectors 
\cite{Cyr-Racine:2013jua,Lancaster:2017ksf,Buen-Abad:2017gxg,DiValentino:2017oaw,DEramo:2018vss,Poulin:2018zxs,Kreisch:2019yzn,Blinov:2019gcj}. 
Another promising early-time solution, is an exotic early dark energy (EDE) component that 
remains subdominant for the majority of the cosmological evolution of the Universe and injects 
a small amount of energy in a very narrow redshift window 
\cite{Poulin:2018cxd,Agrawal:2019lmo,Smith:2019ihp,Alexander:2019rsc,Lin:2019qug,Braglia:2020bym}. 
Note that features in the primordial power spectrum \cite{Hazra:2018opk,Liu:2019dxr} 
and modifications of the recombination history \cite{Chiang:2018xpn,Liu:2019awo} are not able to increase $H_0$.

Another interesting possibility is to alleviate the $H_0$ {\em tension} through a modification 
of General Relativity (GR) \cite{Umilta:2015cta,Ballardini:2016cvy,Nunes:2018xbm,Lin:2018nxe,Rossi:2019lgt,Sola:2019jek,Zumalacarregui:2020cjh,Wang:2020zfv,Ballesteros:2020sik,Braglia:2020iik}, 
which can include early- and late-time modifications of the expansion history with respect 
to $\Lambda$CDM. Scalar-tensor theories of gravity that involve a scalar field non-minimally 
coupled to the Ricci scalar naturally change the effective relativistic degrees of freedom in 
the radiation-dominated epoch and the background expansion history from $\Lambda$CDM, thus 
leading to a higher CMB-inferred $H_0$ \cite{Umilta:2015cta,Ballardini:2016cvy,Rossi:2019lgt}. 
This has been shown for the extended Jordan-Brans-Dicke (eJBD) model in 
Refs.~\cite{Umilta:2015cta,Ballardini:2016cvy} 
where the cosmological parameters estimation has been carried out using data from the 
{\em Planck} 2013 and 2015 release respectively. The robustness of a higher $H_0$ in these 
theories has been proved in Ref.~\cite{Rossi:2019lgt}, where a more general form of the non-minimal 
coupling (NMC) to gravity has been considered. In this paper, we confirm earlier results for 
the particular cases of eJBD and a conformally coupled (CC) scalar field using the most recent 
cosmological data.

Since the mechanism that drives a higher inferred $H_0$ relies on the radiation-like behavior
of the scalar field at early times, it is interesting to investigate to what extent these 
simple scalar-tensor theories are degenerate with effective number of relativistic 
species due to neutrinos and to any additional massless particles 
produced well before recombination $N_{\rm eff}$. 
The current tight constraints from the latest {\em Planck} 2018 data 
$N_{\rm eff} = 2.89 \pm 0.19$ ($N_{\rm eff} = 2.99 \pm 0.17$ including BAO) at 68\% CL 
\cite{Aghanim:2018eyx} can be changed in modified gravity theories as previously shown in 
the context of $f(R)$ gravity in~\cite{Motohashi:2012wc,Chudaykin:2014oia}.

While changing $N_{\rm eff}$ can lead to a higher value for $H_0$ compared with the value 
inferred in the $\Lambda$CDM model from the CMB anisotropies measurements,  
in the extension of $\Lambda$CDM model with non-zero total neutrino mass $m_\nu$, 
lower values of $H_0$ correspond to higher values of $m_\nu$ and higher values of $H_0$ 
correspond to lower $m_\nu$.
For instance, the constraint from {\em Planck} 2018 data in combination with BAO data 
is $m_\nu < 0.12$ eV at 95\% CL \cite{Aghanim:2018eyx,Vagnozzi:2017ovm} and combining 
with the measurement from the SH0ES team the limit further tightens to $m_\nu < 0.076$ eV 
\cite{RoyChoudhury:2019hls}.
In scalar-tensor theories of gravity, it is possible to keep fixed the angular 
diameter distance at decoupling or even increase it in order to recover a higher 
$H_0$ while increasing the total neutrino mass value \cite{Barreira:2014ija}.
Given that neutrino oscillations are the only laboratory evidence of physics beyond the Standard 
Model of Particle Physics \cite{Tanabashi:2018oca}, it is natural to ask ourselves either how the 
cosmological constraints on neutrino physics can be relaxed in these simple models of scalar-tensor 
theories compared to the $\Lambda$CDM concordance model and how much constraints become larger 
in the NMC models including the neutrino parameters. 

Future CMB experiments, such as the Simons Observatory\footnote{\href{https://simonsobservatory.org/}{https://simonsobservatory.org/}} 
\cite{Ade:2018sbj}, 
CMB-S4\footnote{\href{https://cmb-s4.org/}{https://cmb-s4.org/}} \cite{Abazajian:2016yjj},
and future LSS surveys from 
DESI\footnote{\href{http://desi.lbl.gov/}{http://desi.lbl.gov/}} \cite{Levi:2013gra}, 
Euclid\footnote{\href{http://sci.esa.int/euclid/}{http://sci.esa.int/euclid/}} 
\cite{Laureijs:2011gra,Amendola:2012ys}, 
LSST\footnote{\href{http://www.lsst.org/}{http://www.lsst.org/}} 
\cite{Abell:2009aa}, 
SKA\footnote{\href{http://www.skatelescope.org/}{http://www.skatelescope.org/}} 
\cite{Maartens:2015mra,Bacon:2018dui} will help to improve the constraints on these 
extended cosmologies \cite{Brinckmann:2018owf,Alonso:2016suf,Ballardini:2019tho}
and limit the degeneracy of neutrino parameters $N_{\rm eff}$ and $m_\nu$ 
with scalar-tensor theories \cite{Bellomo:2016xhl}.

The paper is organized as follows. Sec.~\ref{sec:models} is devoted to a description 
of the models considered. We describe the datasets considered in Sec.~\ref{sec:data}.
In in Sec.~\ref{sec:update}, we start by discussing the updated constraints on the 
eJBD and CC model in light of new CMB and BAO data,
then we discuss the combination with the distance-ladder measure of $H_0$ from 
SNe Ia in Sec.~\ref{sec:H0}.
We further improve the analysis by extending the cosmological 
parameters to the ones describing the neutrino sector in Secs.~\ref{sec:Neff}-\ref{sec:mnu}-\ref{sec:joint}. 
We draw our conclusions in Sec.~\ref{sec:conclusion}. 
In App.~\ref{sec:appendix}, we collect all the tables with the constraints on the cosmological 
parameters obtained with our MCMC analysis.

\section{The model: non-minimally coupled scalar-tensor theory} \label{sec:models}

As far as cosmological tests are concerned, one of workhorse models to test deviations from GR is 
the eJBD \cite{Jordan:1949zz,Brans:1961sx} theory, which has been extensively studied 
\cite{Chen:1999qh,Nagata:2003qn,Acquaviva:2004ti,Avilez:2013dxa,Li:2013nwa,Ooba:2016slp,Umilta:2015cta,Ballardini:2016cvy,Sola:2019jek}. 
NMC eJBD theory is perhaps the simplest extension to GR within the more general Horndeski theory \cite{Horndeski:1974wa}:
\begin{equation}
    S = \int \dd^{4}x \sqrt{-g} \left[ \frac{F(\sigma)}{2}R 
    - \frac{g^{\mu\nu}}{2} \partial_\mu \sigma \partial_\nu \sigma - V(\sigma) + {\cal L}_m \right] \,,
\end{equation}
where $F(\sigma) = N_{pl}^2+\xi\sigma^2$, $R$ is the Ricci scalar, and ${\cal L}_m$ is the
Lagrangian density for matter fields. As in \cite{Rossi:2019lgt}, we do not consider a quintessence-like 
inverse power-law potential (see for instance 
\cite{Uzan:1999ch,Perrotta:1999am,Bartolo:1999sq,Amendola:1999qq,Chiba:1999wt}), but we restrict ourselves 
to a potential of the type $V(\sigma) = \lambda F^2(\sigma)/4$ for which the scalar field is effectively 
massless. Our choice reduces to eJBD theory after the redefinition 
$\sigma^2 = \phi/(8\pi\xi)$, $\xi = 1/(4 \omega_{\rm BD})$, and setting $N_{pl}=0$.

For a flat Friedmann-Lema{\^ i}tre-Robertson-Walker (FLRW) Universe with $\dd s^2 = - \dd t^2 + a^2(t)\dd {\bf x}^2$, 
the background Friedmann equations in the Jordan frame are \cite{Boisseau:2000pr}:
\begin{align}
    &3FH^2 = \rho_m + \frac{\dot{\sigma}^2}{2} + V(\sigma) - 3 H \dot{F} \,,\\
    &-2F\dot{H} = \rho_m + \dot{\sigma}^2 + \Ddot{F} - H\dot{F} \,.
\end{align}
These equations are supplemented by the Klein-Gordon equation that governs the evolution of the scalar field:
\begin{equation} \label{eq:KG}
    \square\sigma - V_\sigma + F_\sigma R = 0 \,.
\end{equation}
The coupling between gravity and the scalar degree of freedom induces a time varying Newton's 
gravitational constant $G_N$, which is given by $G_N = 1/(8\pi F)$. This quantity  is usually 
denoted by the cosmological Newton's gravitational constant, as opposed to the one that is actually 
measured in laboratory Cavendish-type experiments which is rather given by \cite{Boisseau:2000pr}:
\begin{equation}
    G_{\rm eff} = \frac{1}{8\pi F}\frac{2F+4F^2_{,\sigma}}{2F+3F^2_{,\sigma}} \,.
\end{equation}

Deviations from GR for a theory of gravitation are parameterized by the so called post-Newtonian (PN) 
expansion of the metric \cite{Will:2014kxa}. In such an expansion, the line element can be expressed as:
\begin{equation}
    \dd s^2=-(1+2\Phi-2\beta_{\rm PN}\Phi^2)\dd t^2+(1-2\gamma_{\rm PN}\Phi)\dd {\bf x}^2 \,,
\end{equation}
where we have retained only the two non-null contributions to the PN expansion in the 
case of NMC theories, that is \cite{Boisseau:2000pr}
\begin{equation} \label{eqn:PPN}
    \gamma_{\rm PN}=1-\frac{F_{,\sigma}^{2}}{F+2F_{,\sigma}^{2}} \,,\qquad
    \beta_{\rm PN}=1+\frac{FF_{,\sigma}}{8F+12F_{,\sigma}^{2}}\frac{\dd\gamma_{\rm PN}}{\dd\sigma} \,.
\end{equation}
Solar-system experiments agree with GR predictions, for which both $\gamma_{\rm PN}$ and 
$\beta_{\rm PN}$  are identically equal to unity, at a very precise level. Measurements of 
the perihelion shift of Mercury constrain $\beta_{\rm PN}-1 = (4.1\pm7.8) \times 10^{-5}$ at 68\% CL 
\cite{Will:2014kxa} and Shapiro time delay constrains 
$\gamma_{\rm PN}-1 = (2.1 \pm 2.3) \times 10^{-5}$ at 68\% CL \cite{Bertotti:2003rm}.

In this paper we restrict ourselves to two simple models, both of which only contain one 
extra parameter  with respect to the $\Lambda$CDM model: induced gravity (IG) described by 
$N_{pl} = 0$ and $\xi > 0$, and a conformally coupled scalar field (CC) for which  $\xi = -1/6$ 
and the free parameter is $N_{pl}>M_{pl}$.
For both models, the effective value of the Newton's gravitational constant $G_N$ decreases 
with time, whereas the scalar field $\sigma$ increases for $\xi > 0$, e.g. for IG, and it 
decreases for $\xi < 0$, e.g. for CC\footnote{We refer the interested reader to 
Refs.~\cite{Umilta:2015cta,Ballardini:2016cvy,Rossi:2019lgt} for a detailed analysis of the 
background dynamics in these models, see also Ref.~\cite{Kamenshchik:2017ojc} for the generalization 
to anisotropic homogeneous cosmological models of the Bianchi I type.}. We have $\gamma_{\rm PN} < 1$ 
and $\beta_{\rm PN} \leq 1$ for $\xi > 0$ ($\beta_{\rm PN} = 1$ for IG), whereas $\gamma_{\rm PN} < 1$ 
and $\beta_{\rm PN} > 1$ for $\xi < 0$. 

In order to connect the present value of the field $\sigma_0 \equiv \sigma(z=0)$ to the
gravitational constant, we impose the condition $G_{\rm eff}(\sigma_0) = G$, where 
$G = 6.67 \times 10^{-8}$ cm$^3$ g$^{-1}$ s$^{-2}$ is the gravitational constant measured 
in a Cavendish-type experiment. This corresponds to
\begin{equation} \label{eqn:boundaryIG}
    \tilde{\sigma}_0^2 = \frac{1+8\xi}{\xi(1+6\xi)}
\end{equation}
for IG and
\begin{equation} \label{eqn:boundaryCC}
    \tilde{\sigma}_0^2 = \frac{18 \tilde{N}_{pl}^2 (\tilde{N}_{pl}^2 - 1)}{1 + 3\tilde{N}_{pl}^2} \,,
\end{equation}
for the CC case.
For simplicity, we denote  quantities normalized to $M_{pl} \equiv 1/\sqrt{8\pi G}$ with a tilde.

\section{Methodology and datasets}
\label{sec:data}

In order to derive the constraints on the cosmological parameters we use 
a Markov Chain Monte Carlo (MCMC) 
analysis by using the publicly available code {\tt MontePython}\footnote{\href{https://github.com/brinckmann/montepython\_public}{https://github.com/brinckmann/montepython\_public}} 
\cite{Audren:2012wb,Brinckmann:2018cvx} connected to our modified version of the code 
{\tt CLASS}\footnote{\href{https://github.com/lesgourg/class\_public}{https://github.com/lesgourg/class\_public}} 
\cite{Lesgourgues:2011re,Blas:2011rf}, i.e. {\tt CLASSig} \cite{Umilta:2015cta}. 
Mean values and uncertainties on the parameters  reported, as well as the contours plotted, 
have been obtained using {\tt GetDist}\footnote{\href{https://getdist.readthedocs.io/en/latest}{https://getdist.readthedocs.io/en/latest}} 
\cite{Lewis:2019xzd}.
We use adiabatic initial conditions for the scalar field perturbations \cite{Rossi:2019lgt,Paoletti:2018xet}.

As baseline, we vary the six cosmological parameters for a flat $\Lambda$CDM concordance model, i.e. 
$\omega_b$, $\omega_c$, $H_0$, $\tau$, $\ln\left(10^{10}A_s\right)$, $n_s$, plus one extra parameter 
related to the coupling to the Ricci curvature. 
For IG $(N_{pl}=0,\,\xi>0)$, we sample on the quantity $\zeta_{\rm IG} \equiv \ln\left(1 + 4\xi\right)$, 
according to \cite{Li:2013nwa,Umilta:2015cta,Ballardini:2016cvy} in the prior range $[0,\,0.039]$. 
For CC $(N_{pl}>M_{pl},\,\xi=-1/6)$, we sample on $\Delta \tilde{N}_{pl} \equiv \tilde{N}_{pl} - 1$, 
according to \cite{Rossi:2019lgt}, with prior range $[0,\,0.5]$.
In our updated analysis in Sec.~\ref{sec:update} we assume 3 massless neutrino with $N_{\rm eff} = 3.046$, 
but a more general neutrino sector is considered in the rest of the paper. 

We quote constraints on the variation of the Newton's gravitational constant between the 
radiation era and the present time $\delta G_N/G_N$, and its derivative at present time $\dot{G}_N/G_N$.
Defining the effective cosmological gravitational strength \cite{Avilez:2013dxa}:
\begin{equation}
    \frac{G_N}{G}(z=0) = \frac{1}{F(\sigma_0)} \,,
\end{equation}
we have 
\begin{align}
    &\frac{\delta G_N}{G_N}(z=0) = \frac{G_N(\sigma_0)-G_N(\sigma_{\rm ini})}{G_N(\sigma_{\rm ini})} 
    = \frac{\sigma_{\rm ini}^2-\sigma_0^2}{\sigma_0^2} \leq 0 \label{eqn:ig_deltaG} \,,\\
    &\frac{\dot{G}_N}{G_N}(z=0) = -\frac{2\dot{\sigma}_0}{\sigma_0} \leq 0 \label{eqn:ig_dotG} \,,\\
    &\frac{G_N}{G}(z=0) = \frac{1+6\xi}{1+8\xi} \leq 1 \,,
\end{align}
for IG and 
\begin{align}
    &\frac{\delta G_N}{G_N}(z=0) = \xi\frac{\sigma_{\rm ini}^2-\sigma_0^2}{N_{pl}^2+\xi\sigma_0^2} 
    = \frac{\sigma_0^2-\sigma_{\rm ini}^2}{6N_{pl}^2-\sigma_0^2} \leq 0 \label{eqn:cc_deltaG} \,,\\
    &\frac{\dot{G}_N}{G_N}(z=0) = -\frac{2\xi\dot{\sigma}_0\sigma_0}{N_{pl}^2+\xi\sigma_0^2} 
    = \frac{2\dot{\sigma}_0\sigma_0}{6N_{pl}^2-\sigma_0^2} \leq 0 \label{eqn:cc_dotG} \,,\\
    &\frac{G_N}{G}(z=0) = \frac{1}{4}\left(3+\frac{M_{pl}^2}{N_{pl}^2}\right) \leq 1 \,,
\end{align}
for the CC case, where $\delta G_N/G_N,\, \dot{G}_N/G_N = 0$ and $G_N/G = 1$ correspond to 
the predictions in GR.

We also list the $\Delta \chi^2 \equiv \chi^2 - \chi^2_{\Lambda CDM}$, where negative values indicate 
a better-fit to the datasets with respect to the $\Lambda$CDM model.

\subsection{Datasets}

We constrain the cosmological parameters using several combination of datasets. 
Our CMB measurements are those from the {\em Planck} 2018 legacy release (hereafter P18) 
on temperature, polarization, and weak lensing CMB anisotropies angular power spectra 
\cite{Aghanim:2019ame,Aghanim:2018oex}. 
The high-multipoles likelihood $\ell \geq 30$ is based on {\tt Plik} likelihood. 
We use the low-$\ell$ likelihood combination at $2 \leq \ell < 30$: temperature-only 
{\tt Commander} likelihood plus the {\tt SimAll} EE-only likelihood.
For the {\em Planck} CMB lensing likelihood, we consider the {\em conservative} 
multipoles range, i.e. $8 \leq \ell \leq 400$. We marginalize over foreground and 
calibration nuisance parameters of the {\em Planck} likelihoods \cite{Aghanim:2019ame,Aghanim:2018oex} 
which are also varied together with the cosmological ones.

Baryon acoustic oscillation (BAO) measurements from galaxy redshift surveys are used 
as primary astrophysical dataset to constraint these class of theories providing a 
complementary late-time information to the CMB anisotropies. We use the 
Baryon Spectroscopic Survey (BOSS) DR12 \cite{Alam:2016hwk} {\em consensus} results on 
BAOs in three redshift slices with effective redshifts $z_{\rm eff} = 0.38,\,0.51,\,0.61$ 
\cite{Ross:2016gvb,Vargas-Magana:2016imr,Beutler:2016ixs}, 
in combination with measure from 6dF \cite{Beutler:2011hx} at $z_{\rm eff} = 0.106$ 
and the one from SDSS DR7 \cite{Ross:2014qpa} at $z_{\rm eff} = 0.15$.

Finally, we consider the combination with a Gaussian likelihood based on the 
determination of the Hubble constant from from Hubble Space Telescope (HST) observations 
(hereafter R19), i.e. $H_0 = (74.03 \pm 1.42)$ km s$^{-1}$Mpc$^{-1}$ \cite{Riess:2019cxk}.

\section{Updated {\em Planck} 2018 results} \label{sec:update}

The results in this section update those obtained in Ref.~\cite{Ballardini:2016cvy} 
for IG and in Ref.~\cite{Rossi:2019lgt} for CC, based on the {\em Planck} 2015 data (P15) 
\cite{Adam:2015rua,Aghanim:2015xee} in combination with an older compilation of BAO data, 
i.e. DR10-DR11 \cite{Beutler:2011hx,Anderson:2013zyy,Ross:2014qpa}. 

\begin{figure}
\centering
\includegraphics[height=0.45\textwidth]{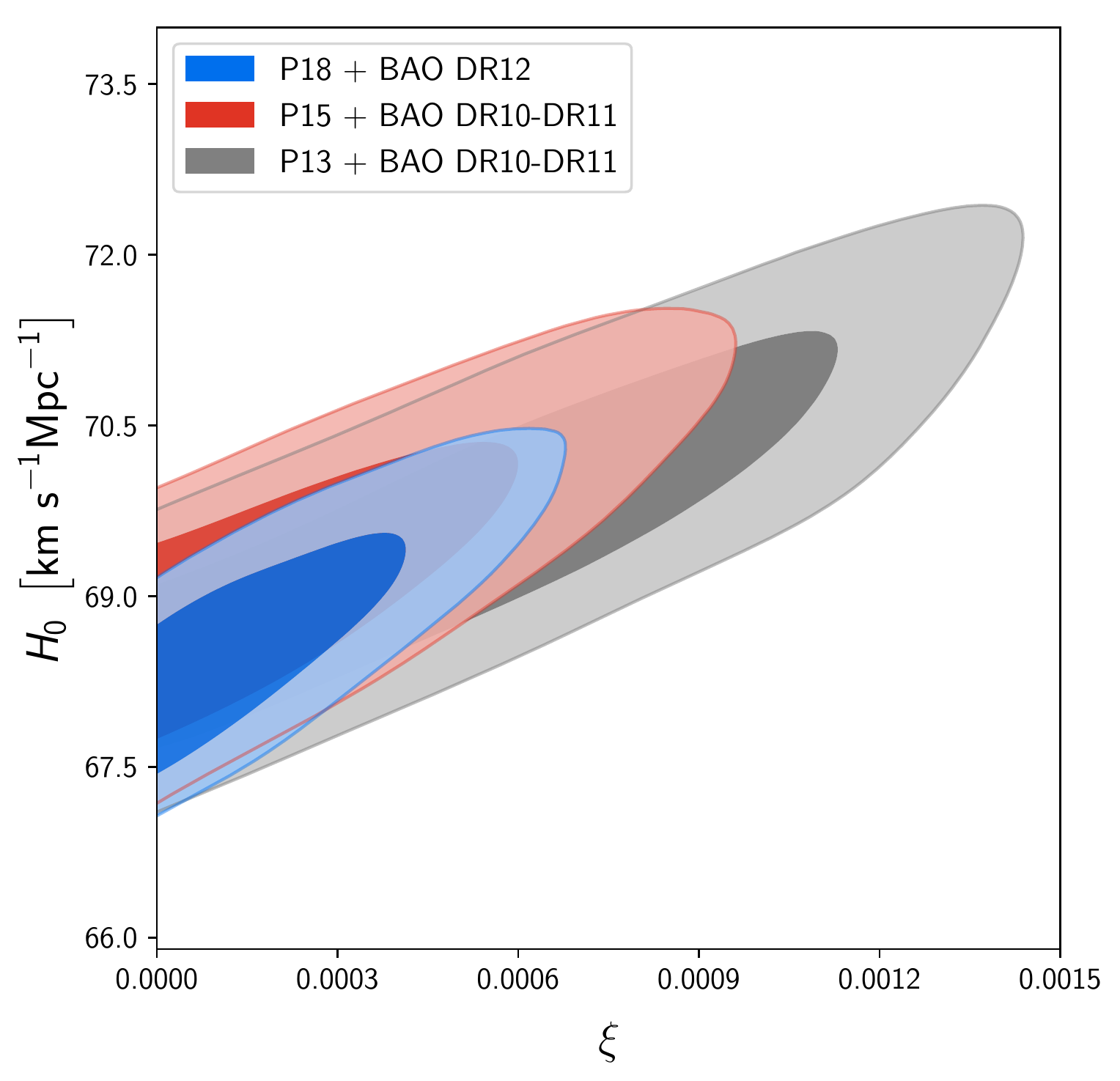}
\includegraphics[height=0.45\textwidth]{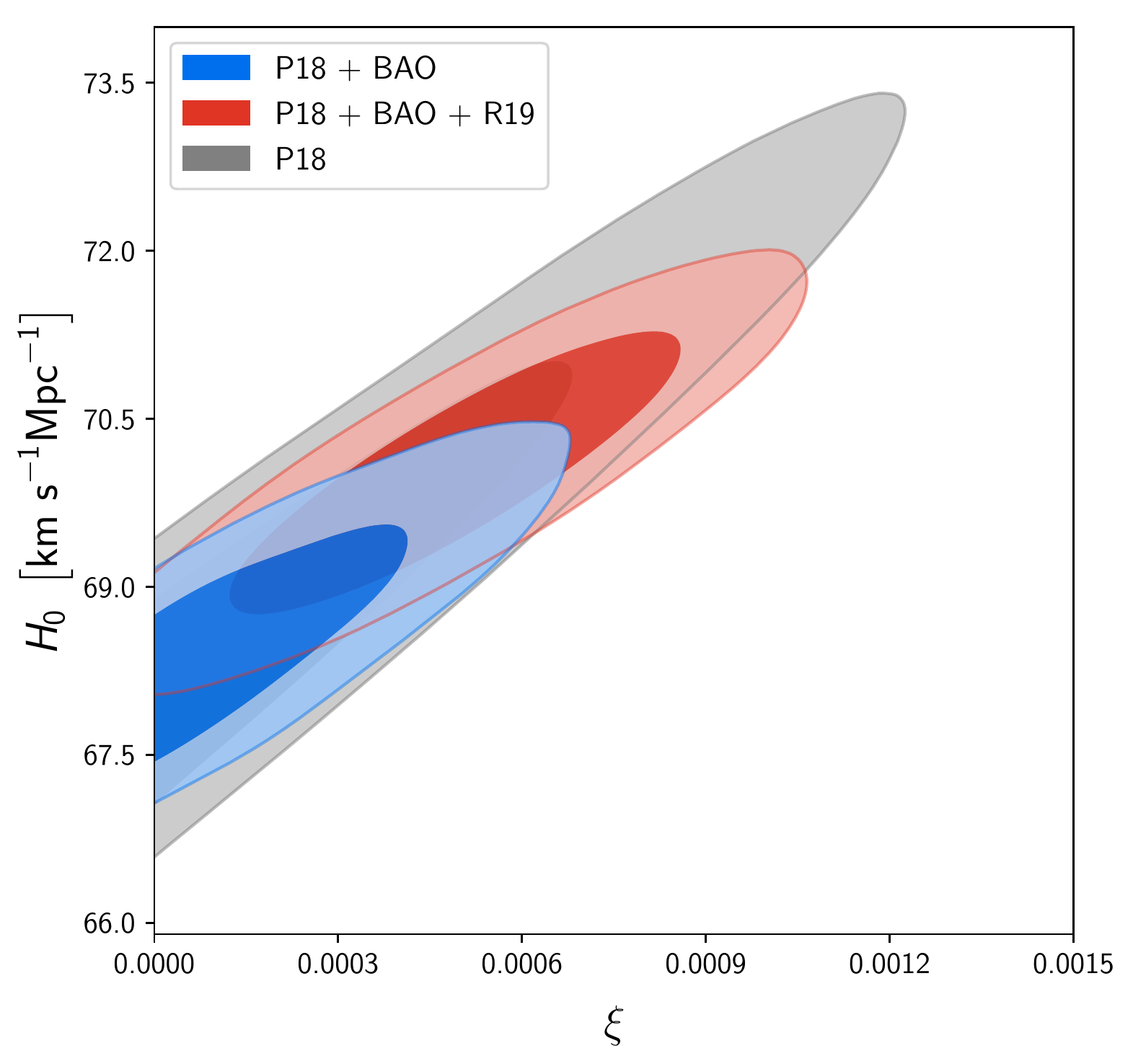}
\caption{Left panel: 
marginalized joint 68\% and 95\% CL regions 2D parameter space using 
current versus previous releases of {\em Planck} data and BOSS BAO data 
from \cite{Umilta:2015cta,Ballardini:2016cvy}. 
Right panel: marginalized joint 68\% and 95\% CL regions 2D parameter space using 
P18 (gray) in combination with BAO (blue) and BAO + R19 (red) for the IG model.}
\label{fig:ig_H0}
\end{figure}

First, we discuss the IG case. We find that the constraint on the coupling parameter $\xi$ 
obtained from the CMB alone is almost half of the bound obtained with P15 which was 
$\xi < 0.0017$ at 95\% CL. With the full  high-$\ell$ polarization information and the new 
determination of $\tau$ we obtain $\xi < 0.00098$ at 95\% CL. Adding the BAO data, we obtain 
$\xi < 0.00055$ at 95\% CL, which is 25\% tighter compared 
to the limit obtained with P15 in combination with BAO DR10-11, i.e. $\xi < 0.00075$ and 
half of the one obtained with P13 in combination with BAO DR10-11, i.e. $\xi < 0.0012$, 
see the left panel of  Fig.~\ref{fig:ig_H0}.
As we can see from Tab.~\ref{tab:ig}, BAO data strongly constrain the model and are useful 
to break the degeneracy in the $H_0-\xi$ parameter space.

\begin{figure}
\centering
\includegraphics[height=0.45\textwidth]{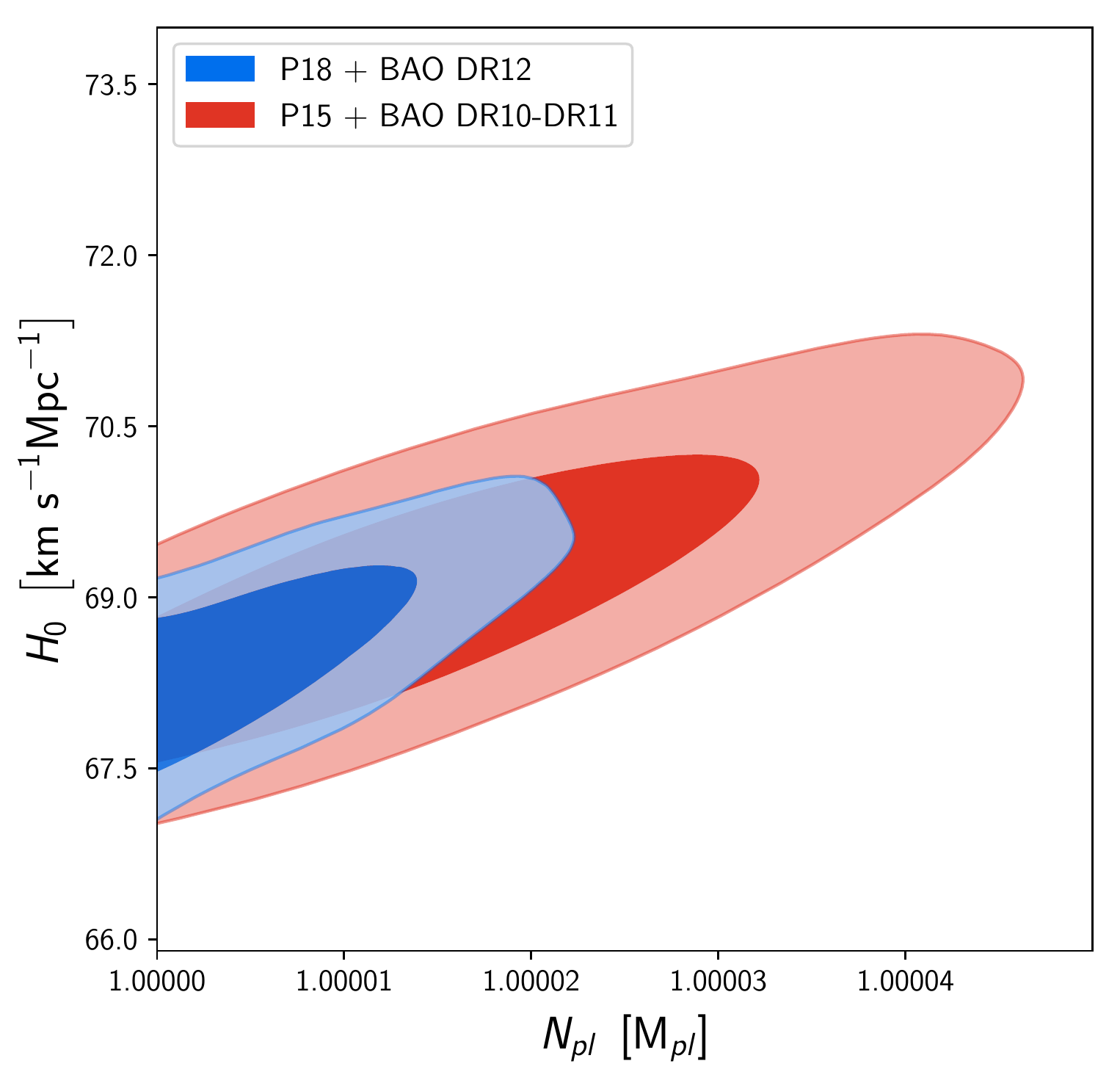}
\includegraphics[height=0.45\textwidth]{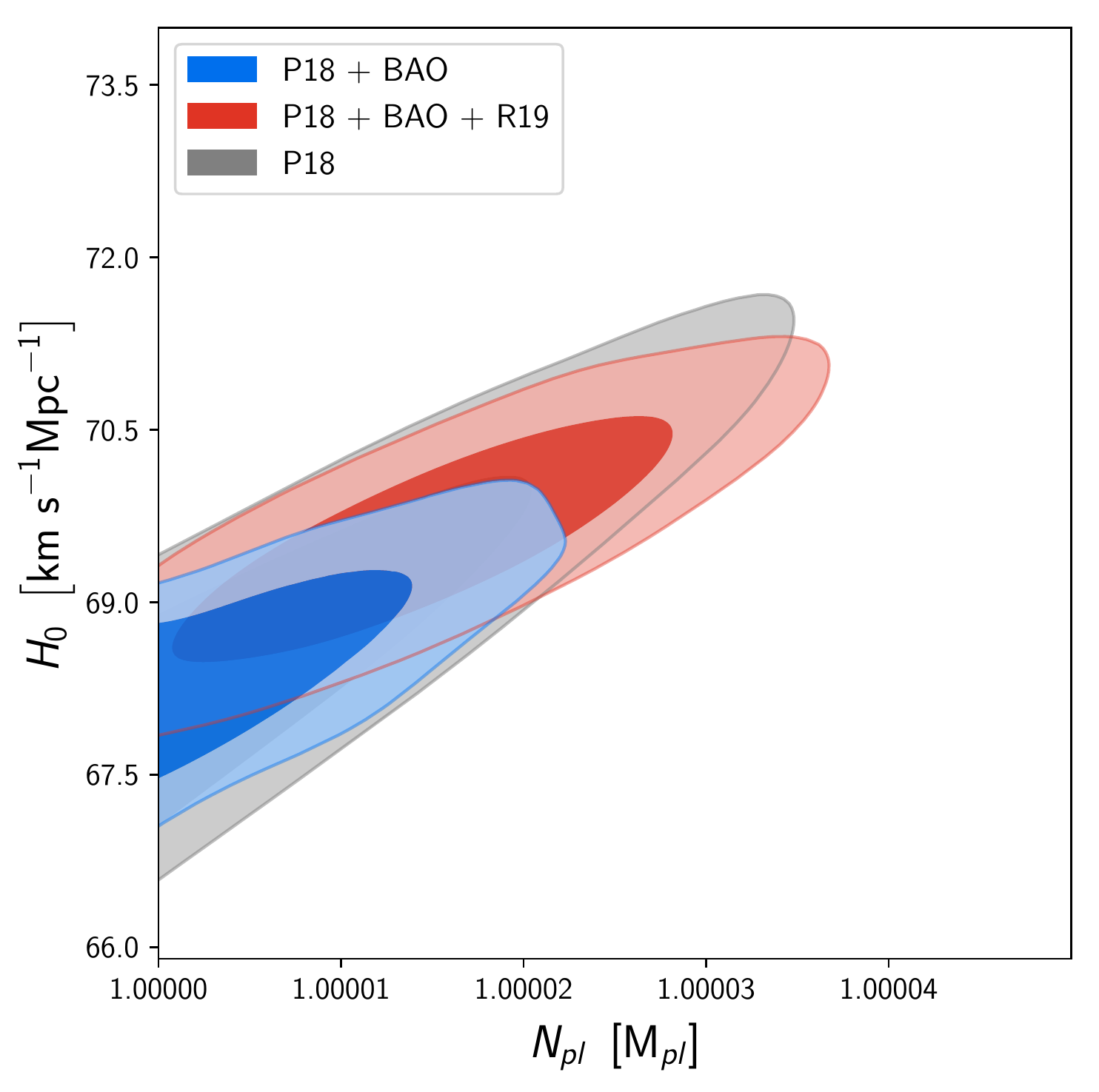}
\caption{Left panel: 
marginalized joint 68\% and 95\% CL regions 2D parameter space using 
P18 (P15) data in combination BAO in blue (red).
Right panel: marginalized joint 68\% and 95\% CL regions 2D parameter space using 
P18 (gray) in combination with BAO (blue) and BAO + R19 (red) for the CC model.}
\label{fig:cc_H0}
\end{figure}

We now discuss  the CC case. The coupling to gravity is constrained to 
$N_{pl} < 1.000028\ $M$_{pl}$ at 95\% CL for P18 and 
$N_{pl} < 1.000018\ $M$_{pl}$ at 95\% CL in combination with BAO data. These constraints 
update the ones obtained with P15 in combination with D10-DR11 BAO 
$N_{pl} < 1.000038\ $M$_{pl}$ at 95\% CL in Ref.~\cite{Rossi:2019lgt}.
As in the IG case, we still have a degeneracy between $H_0$ and the coupling to gravity $N_{pl}$ 
as visible from Fig.~\ref{fig:cc_H0}.
For these NMC models we recover the same cosmological parameters and uncertainties 
if we allow $\xi$ to vary, with prior range $[0,0.1]$ and $[-0.1,0]$, together with $N_{pl}$.
We find for the positive branch ($N_{pl} < M_{pl},\,\xi > 0$) of the coupling:
\begin{equation}
    N_{pl} > 0.64\ {\rm M}_{pl}\ (> 0.60\ {\rm M}_{pl}) \,,\qquad
    \xi < 0.046\ (< 0.055)
\end{equation}
both at 95\% CL and $H_0 = \left(68.78^{+0.56}_{-0.84}\right)$ km s$^{-1}$Mpc$^{-1}$ 
($70.14^{+0.86}_{-0.72}$ km s$^{-1}$Mpc$^{-1}$) with P18+BAO (P18+BAO+R19).
The constraints for the negative branch ($N_{pl} > M_{pl},\,\xi < 0$) are:
\begin{equation}
    N_{pl} < 1.05\ {\rm M}_{pl}\ (< 1.04\ {\rm M}_{pl}) \,,\qquad
    \xi > -0.042\ (> -0.051)
\end{equation}
both at 95\% CL and $H_0 = \left(68.76^{+0.54}_{-0.78}\right)$ km s$^{-1}$Mpc$^{-1}$ 
($69.74 \pm 0.75$ km s$^{-1}$Mpc$^{-1}$) with P18+BAO (P18+BAO+R19).

Consistently with the constraints on the coupling parameters $\xi$ and $N_{pl}$, 
we find also tighter limits on the variation of the Newton's gravitational constant 
\eqref{eqn:ig_deltaG}-\eqref{eqn:cc_deltaG} and its derivative \eqref{eqn:ig_dotG}-\eqref{eqn:cc_dotG} 
at present time. For IG, we have:
\begin{equation}
    \frac{\delta G_{\rm N}}{G_{\rm N}}(z=0) > - 0.016 \,,\qquad \frac{\dot{G}_{\rm N}}{G_{\rm N}}(z=0) > - 0.66\times 10^{-13}\ {\rm yr}^{-1}
\end{equation}
for P18 + BAO at 95\% CL, updating those obtained in Ref.~\cite{Ballardini:2016cvy} 
with P15 + BAO DR10-11, i.e.
\begin{equation}
    \frac{\delta G_{\rm N}}{G_{\rm N}}(z=0) > - 0.027 \,,\qquad \frac{\dot{G}_{\rm N}}{G_{\rm N}}(z=0) > - 1.4\times 10^{-13}\ {\rm yr}^{-1} \,.
\end{equation}
For CC, we obtain the following 95\% CL bounds for P18 + BAO:
\begin{equation}
    \frac{\delta G_{\rm N}}{G_{\rm N}}(z=0) > - 0.017 \,,\qquad \frac{\dot{G}_{\rm N}}{G_{\rm N}}(z=0) > - 0.25\times 10^{-23}\ {\rm yr}^{-1} \,.
\end{equation}
Note that whereas the constraints on $\delta G_{\rm N}/G_{\rm N}(z=0)$ hardly change for different coupling 
$F(\sigma)$, the limits on $\dot{G}_{\rm N}/G_{\rm N}(z=0)$ strongly depend on the details of the model
\footnote{The same behaviour of the constraints on the variation of the Newton's constant and 
its derivative has been observed changing the potential $V(\sigma)$ for IG, see 
Ref.~\cite{Ballardini:2016cvy}.}, 
but are anyway much tighter than those obtained by the Lunar Laser Ranging experiment \cite{Williams:2004qba}. 
Tab.~\ref{tab:ig} also reports the values of the post-Newtonian parameters as derived parameters from 
our samples: whereas for IG $\gamma_{\rm PN} \lesssim {\cal O} (10^{-3})$, for CC the bounds we derive on 
$\gamma_{\rm PN} \,, \beta_{\rm PN}$ are now tighter than those in the Solar System.

\begin{figure}
\centering
\includegraphics[height=0.45\textwidth]{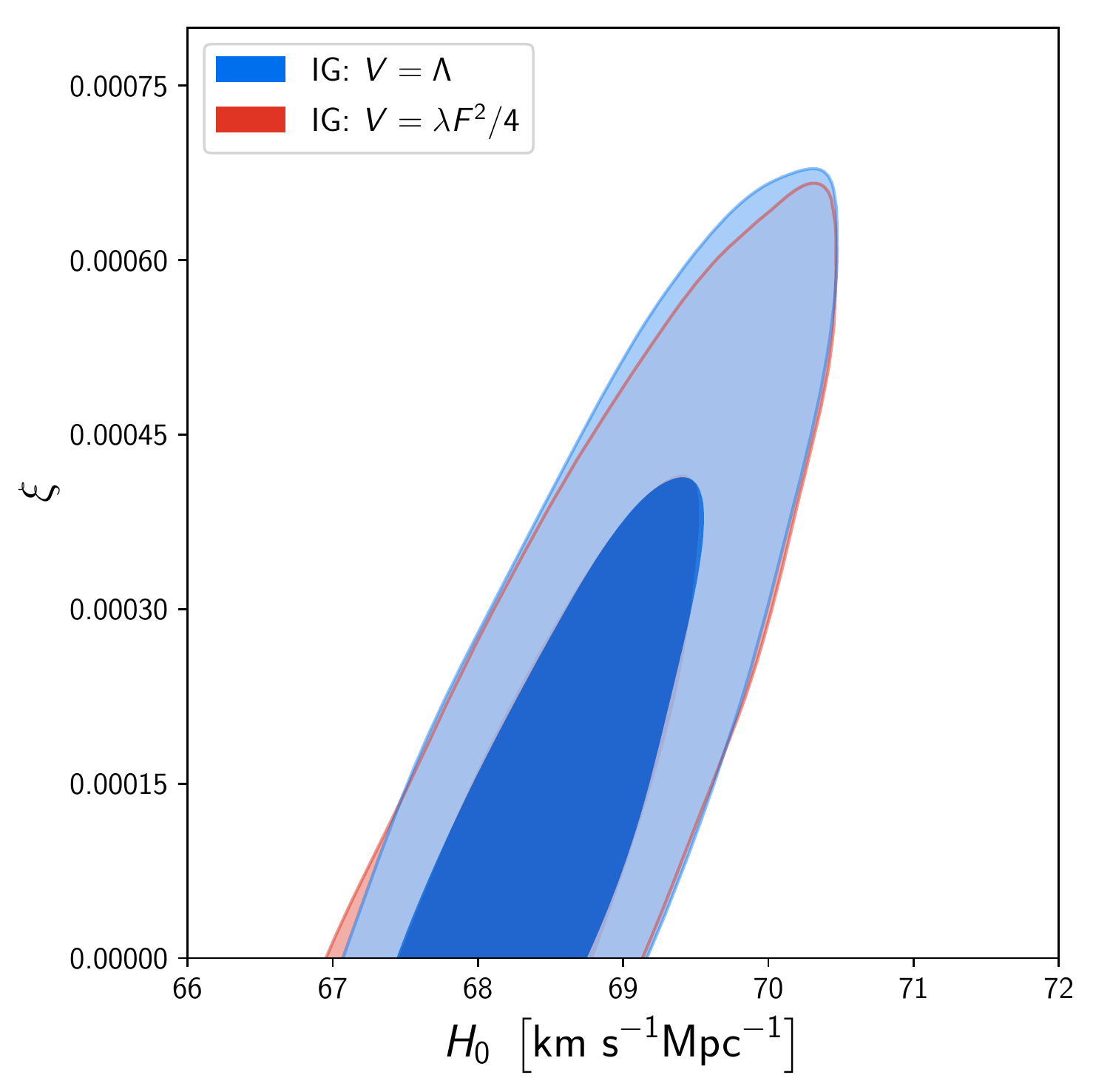}
\includegraphics[height=0.45\textwidth]{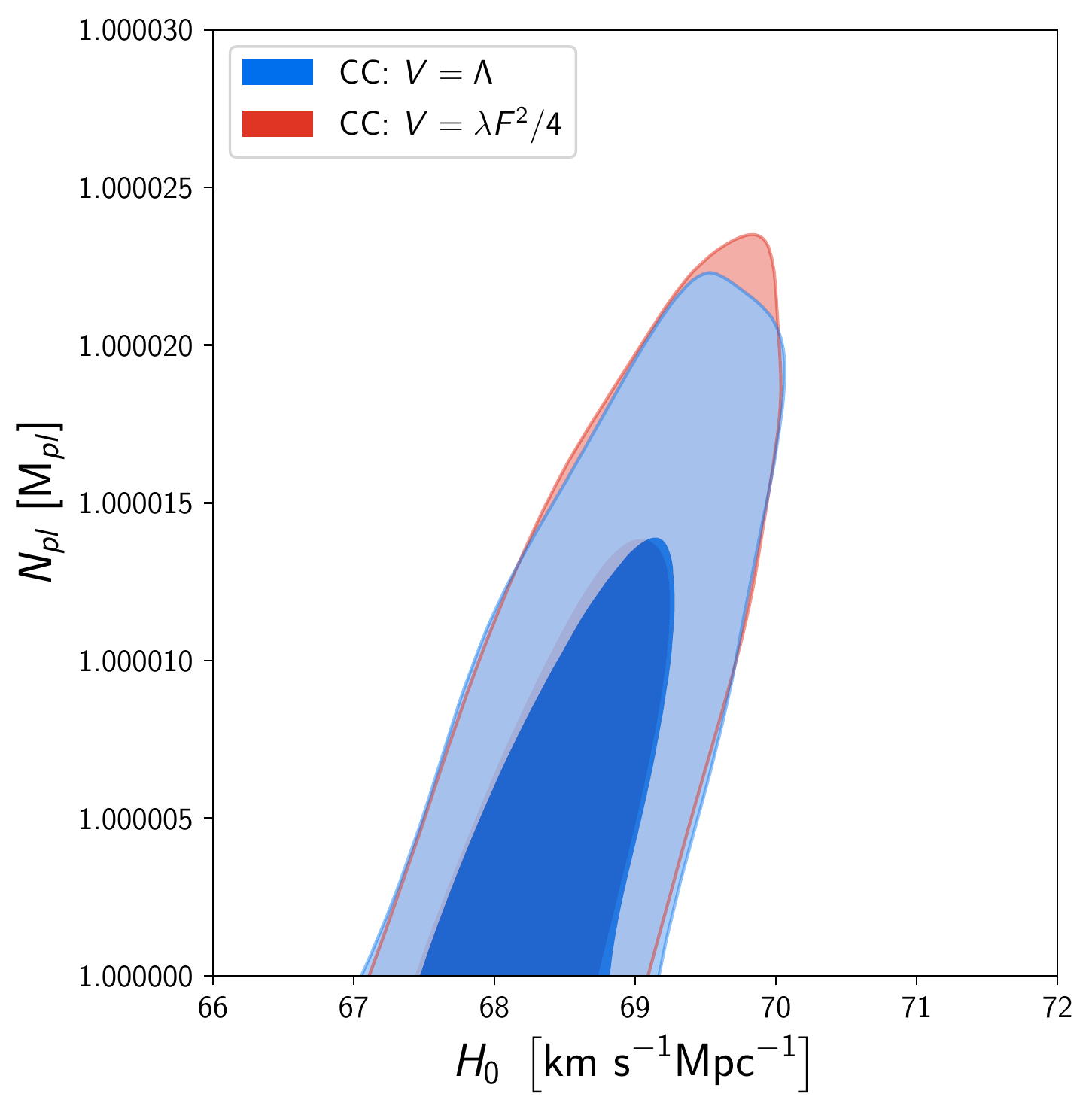}
\caption{Left panel: 
marginalized joint 68\% and 95\% CL regions 2D parameter space $H_0-\xi$ using 
P18 + BAO data for IG with $V(\sigma) = \lambda F(\sigma)^2/4$ (red) 
and $V(\sigma) = \Lambda$ (blue).
Right panel: marginalized joint 68\% and 95\% CL regions 2D parameter space $H_0-N_{pl}$ 
using P18 + BAO data for CC with $V(\sigma) = \lambda F(\sigma)^2/4$ (red) 
and $V(\sigma) = \Lambda$ (blue).}
\label{fig:lambda}
\end{figure}

We note that the value of the initial (final) scalar field is small and sub-Planckian 
$\sigma_{ini} = (0.22 \pm 0.10)\ $M$_{pl}$ ($\sigma_0 = (0.0089 \pm 0.0040)\ $M$_{pl}$) 
as opposed to the IG case where the evolution is super-Planckian.

We have tested that our results are stable when we switch to a flat potential $V(\sigma) = \Lambda$ for 
both IG and CC. We show in Fig.~\ref{fig:lambda} the consistency of the posterior distributions 
for $H_0$, $\xi$, and $N_{pl}$ with current cosmological data (P18 + BAO).
The comparison with $V(\sigma) = \Lambda$ for IG updates the results obtained in \cite{Ballardini:2016cvy} 
with {\em Planck} 2015 which showed that the cosmological parameters were 
stable when varying the index $n$ of a power-law potential $V(\sigma) = \lambda\sigma^n/4$ with $n \ge 0$. 
The stability of the results for the CC case when switching to $V(\sigma) = \Lambda$ is a new result, 
although, in analogy with what happens in IG, not totally unexpected since our data constrains the 
deviations ${\cal O}(\sigma^2/N^2_{pl})$ of $F^2$ from a flat potential.
We refer the interested reader to Ref.~\cite{Braglia:2020iik} for an extended analysis of 
$V(\sigma) = \Lambda$ for $F(\sigma) = M_{pl}^2 \left[1+ \xi (\sigma/M_{pl})^n\right]$ with $n=2,4$ 
with flat priors on 
$\sigma_\mathrm{ini}$\footnote{See also  Ref.~\cite{Ballesteros:2020sik} for similar finding 
in NMC scalar-tensor gravity with a coupling $F(\sigma) = M_{pl}^2 + \xi\sigma^2$ in the parameter 
space range for $\xi < 0$.}.

\section{Implications for the $H_0$ tension and combination with R19 data} \label{sec:H0}

We find a higher value for the Hubble parameter for IG, i.e. $H_0 = \left(69.6^{+0.8}_{-1.7}\right)$ 
km s$^{-1}$Mpc$^{-1}$, and for CC, i.e. $H_0 = \left(69.0^{+0.7}_{-1.2}\right)$ km s$^{-1}$Mpc$^{-1}$, 
compared to the $\Lambda$CDM case, i.e. $H_0 = \left(67.36 \pm 0.54 \right)$ km s$^{-1}$Mpc$^{-1}$, 
for P18.

The addition of BAO drives the value for $H_0$ to lower values, for IG to $H_0 = \left(68.78^{+0.53}_{-0.78}\right)$ 
km s$^{-1}$Mpc$^{-1}$ and for CC to $H_0 = \left(68.62^{+0.47}_{-0.66}\right)$ km s$^{-1}$Mpc$^{-1}$. 
Note however that these values are larger than the corresponding $\Lambda$CDM value, 
i.e. $H_0 = \left(67.66 \pm 0.42 \right)$ km s$^{-1}$Mpc$^{-1}$. 

Once we include R19, we obtain $H_0 = (70.1 \pm 0.8)$ km s$^{-1}$Mpc$^{-1}$ at 68\% CL, 
$\xi = 0.00051^{+0.00043}_{-0.00046}$ at 95\% CL for IG, 
$H_0 = \left(69.64^{+0.65}_{-0.73}\right)$ km s$^{-1}$Mpc$^{-1}$ at 68\% CL, 
$N_{pl} < 1.000031\ $M$_{pl}$ at 95\% CL for CC.
Fig.~\ref{fig:ig_H0} shows how the degeneracy between $H_0$ and $\xi$ 
can easily accommodate for larger $H_0$ value with respect to the $\Lambda$CDM concordance model 
reducing the $H_0$ {\em tension} from 4.4$\sigma$ to 
2.7$\sigma$ (3.2$\sigma$) for P18 and 3.5$\sigma$ (3.6$\sigma$) including BAO for 
IG (CC). The reduction of the tension is 
due to the combination of having an higher mean and larger uncertainties on $H_0$ 
compared to the $\Lambda$CDM model. We find that $H_0$ is about 
$\sim 0.5$ km s$^{-1}$Mpc$^{-1}$ higher in the IG compared to the CC case for every choice 
of datasets combination. 

Note that the models considered in our paper can produce a values of
$H_0$ in complete agreement with the local value of $H_0$ measured using red
giants \cite{Freedman:2019jwv}, though not that measured using SNe Ia \cite{Riess:2019cxk}.

\section{Degeneracy with the neutrino sector} 

\subsection{Effective number of relativistic degrees of freedom}
\label{sec:Neff}

The presence of extra relativistic degrees of freedom in the Universe increases the expansion 
rate during the radiation-dominated era and shifts the epoch of matter-radiation equality, the 
shape of the matter power spectrum, and the history of recombination (see 
Refs.~\cite{Hannestad:2006zg,Lesgourgues:2018ncw} for a review). 
The extra radiation is usually parameterized by $\Delta N_{\rm eff} \equiv N_{\rm eff}-3.046$ 
which takes into account that neutrino decoupling was not quite complete when $e^+e^-$ annihilation 
began \cite{Dodelson:1992km,Fields:1992zb,Dolgov:1992qg,Mangano:2005cc}. 

\begin{figure}
\centering
\includegraphics[width=1.\textwidth]{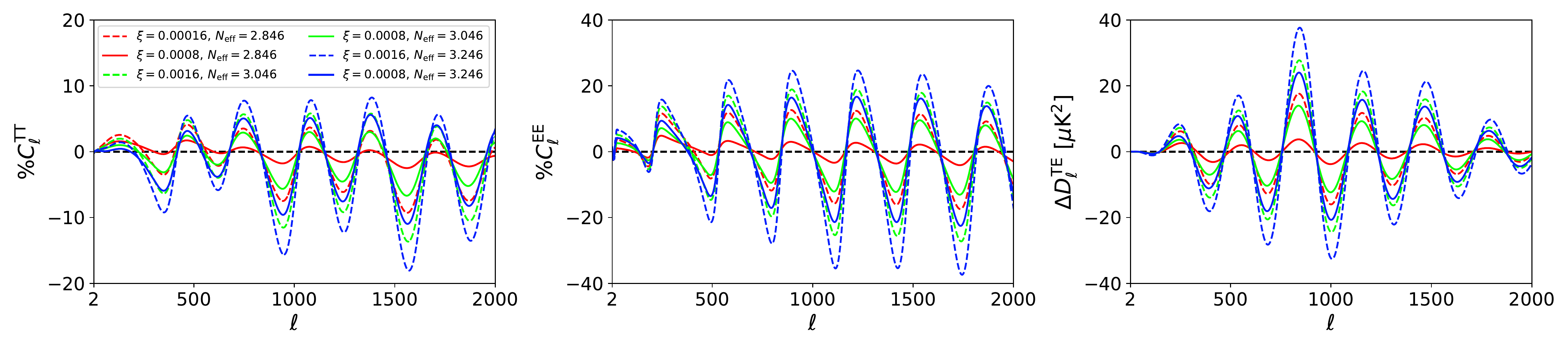}
\includegraphics[width=1.\textwidth]{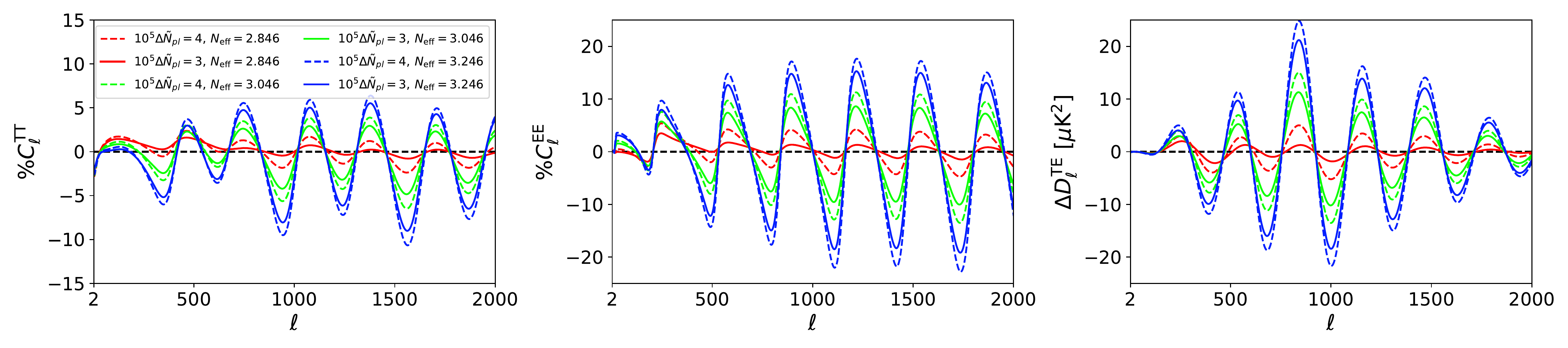}
\caption{Differences with respect to the $\Lambda$CDM with ($N_{\rm eff} = 3.046$) 
with IG (top panels) for $\xi = 0.0008,\,0.0016$ (solid, dashed) and 
$N_{\rm eff} = 2.846,\,3.046,\,3.246$ (red, green, blue), and CC (bottom panels) 
for $N_{pl} = 1.00003,\,1.00004$ M$_{pl}$ (solid, dashed) 
and $N_{\rm eff} = 2.846,\,3.046,\,3.246$ (red, green, blue). 
$D_\ell \equiv \ell(\ell+1)C_\ell/(2\pi)$ are the band-power angular power spectra.}
\label{fig:cl_Neff}
\end{figure}

In the context of the $H_0$ {\em tension}, a larger value of $N_{\rm eff}$ can attenuate the 
discrepancy on the $H_0$ value and reduce the comoving sound horizon at baryon drag. However, 
the tension is only reduced ($\sim 2\sigma$) and it appears again when BAO data is included 
\cite{Bernal:2016gxb,Aghanim:2018eyx}.

\begin{figure}
\centering
\includegraphics[height=0.32\textwidth]{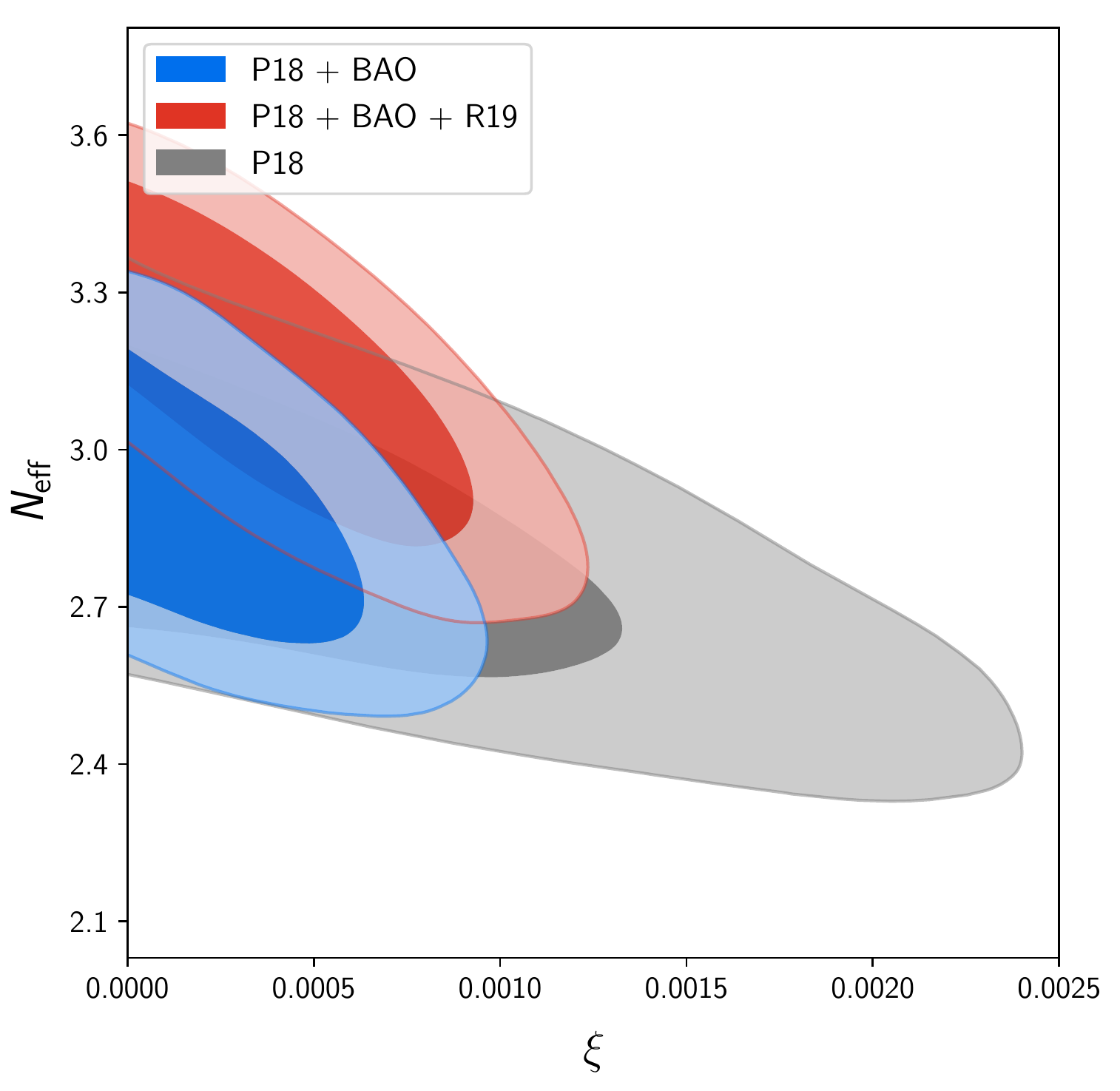}
\includegraphics[height=0.32\textwidth]{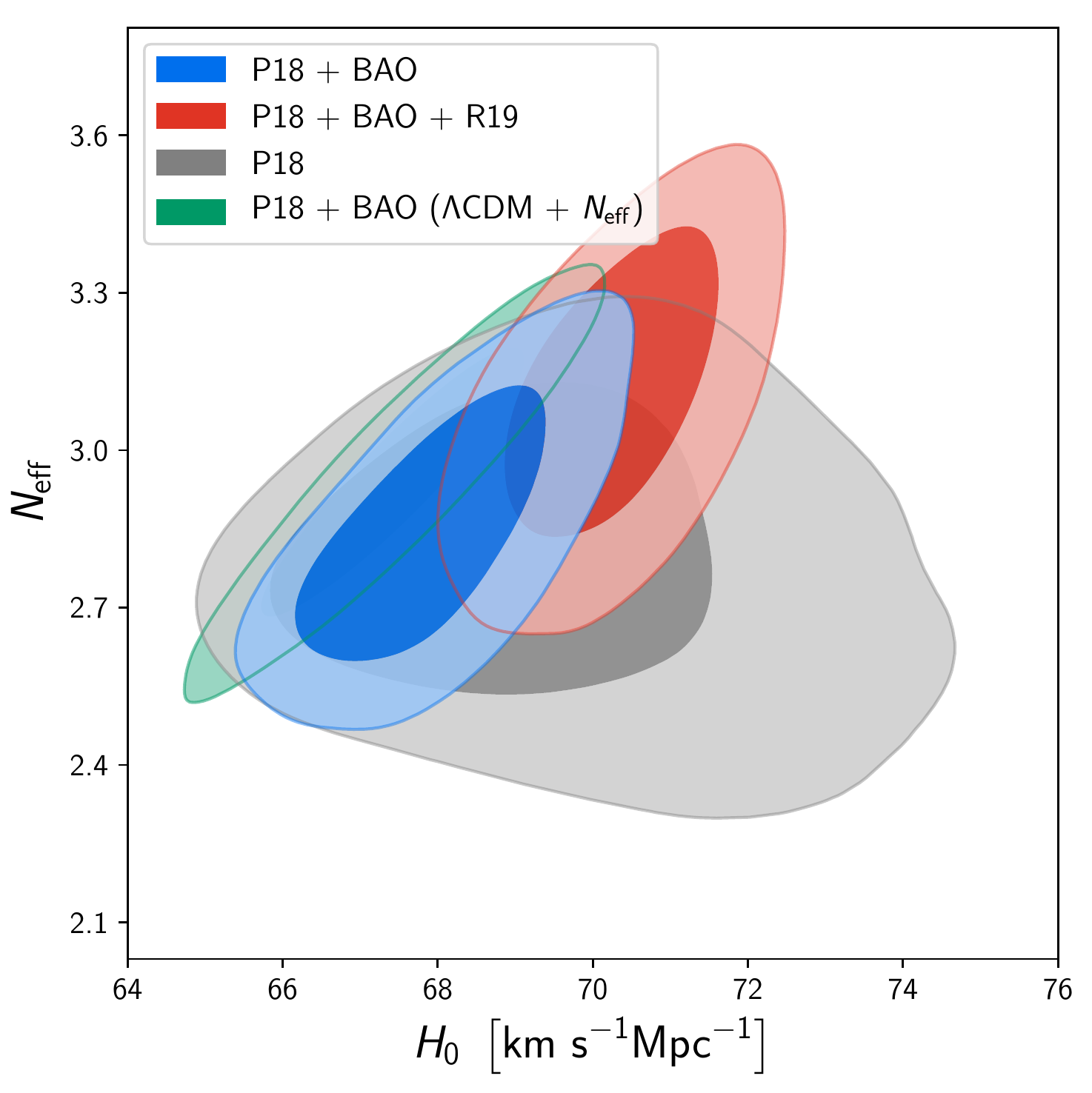}
\includegraphics[height=0.32\textwidth]{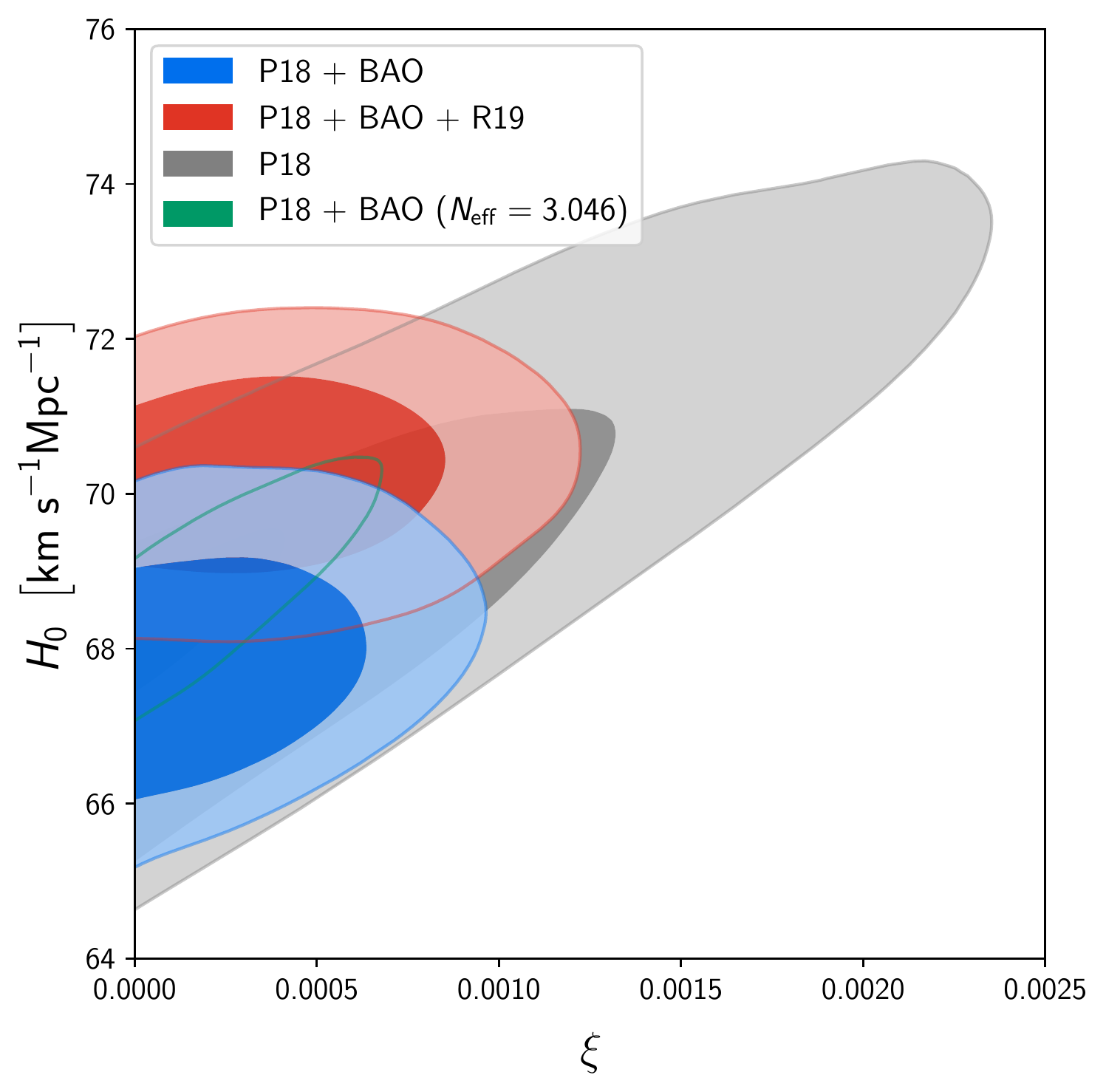}
\caption{Marginalized joint 68\% and 95\% CL regions 2D parameter space using the P18 (gray) 
in combination with BAO (blue) and BAO + R19 (red) for the IG+$N_{\rm eff}$ model. 
In the central panel, we include the $H_0-N_{\rm eff}$ contours for the $\Lambda$CDM in green.
In the right panel, we include the $H_0-\xi$ contours for the IG with $N_{\rm eff} = 3.046$ in green.}
\label{fig:ig_Neff}
\end{figure}

There is an interplay (negative correlation) between the contribution of extra radiation 
with respect to the standard $\Lambda$CDM scenario from the $N_{\rm eff}$ and from the scalar 
field coupling\footnote{See Refs.~\cite{Motohashi:2012wc,Chudaykin:2014oia} for an application 
in the context of $f(R)$ gravity.} which acts as an additional source of radiation in the early Universe. 
Decreasing the effective number of extra relativistic species to $N_{\rm eff} = 2.846$ 
we obtain deviations of the CMB anisotropies angular power spectra to the $\Lambda$CDM 
model of the same order of the ones obtained with  $\xi$ halved and $N_{\rm eff} = 3.046$, 
see Fig.~\ref{fig:cl_Neff}.

We see that, preferring lower values of $N_{\rm eff}$, the dataset allows for larger 
values for $\xi$ compared to the case with $N_{\rm eff} = 3.046$ fixed.
We go from $\xi < 0.00098$ to $\xi < 0.0019$ at 95\% CL with P18 alone and from 
$\xi < 0.00055$ to $\xi < 0.00078$ once we include BAO, see Tab.~\ref{tab:ig_Neff}.

The mean of $N_{\rm eff}$ moves around $1\sigma$ toward lower values with respect to the 
$\Lambda$CDM case with a similar error. For IG, we get at 68\% CL $N_{\rm eff} = 2.79 \pm 0.20$ 
for P18 compared to $N_{\rm eff} = 2.89 \pm 0.19$ in $\Lambda$CDM and $N_{\rm eff} = 2.85 \pm 0.17$ 
in combination with BAO compared to $N_{\rm eff} = 2.99 \pm 0.17$ in $\Lambda$CDM.
In Fig.~\ref{fig:ig_Neff} (central panel), we show the enlarged $H_0-N_{\rm eff}$ parameter 
space in IG compared to the $\Lambda$CDM concordance model (green contours) where it is possible 
to reach higher value of $H_0$ without increasing $N_{\rm eff}$ in presence of a modification 
of gravity.

\begin{figure}
\centering
\includegraphics[width=0.32\textwidth]{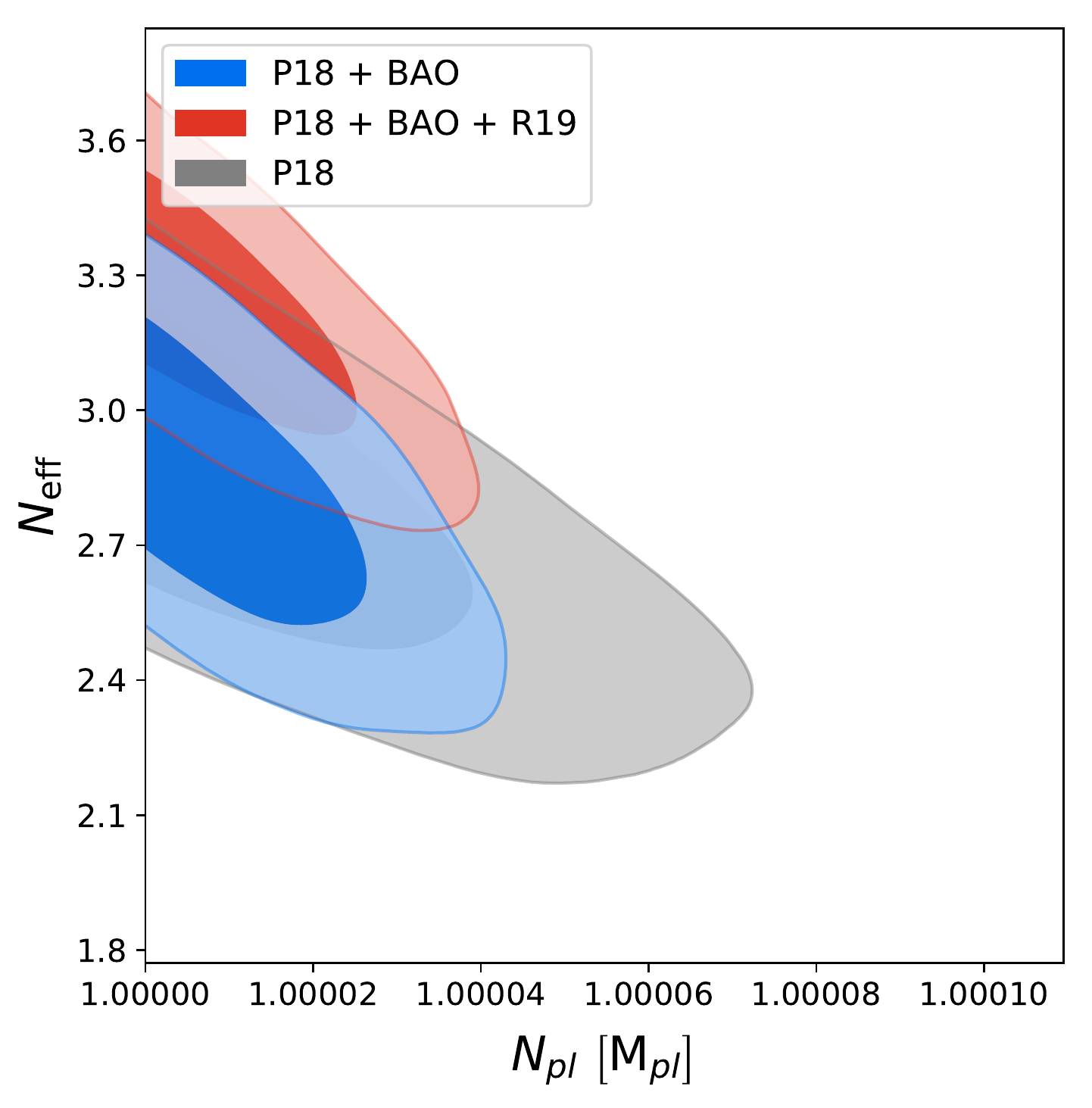}
\includegraphics[width=0.32\textwidth]{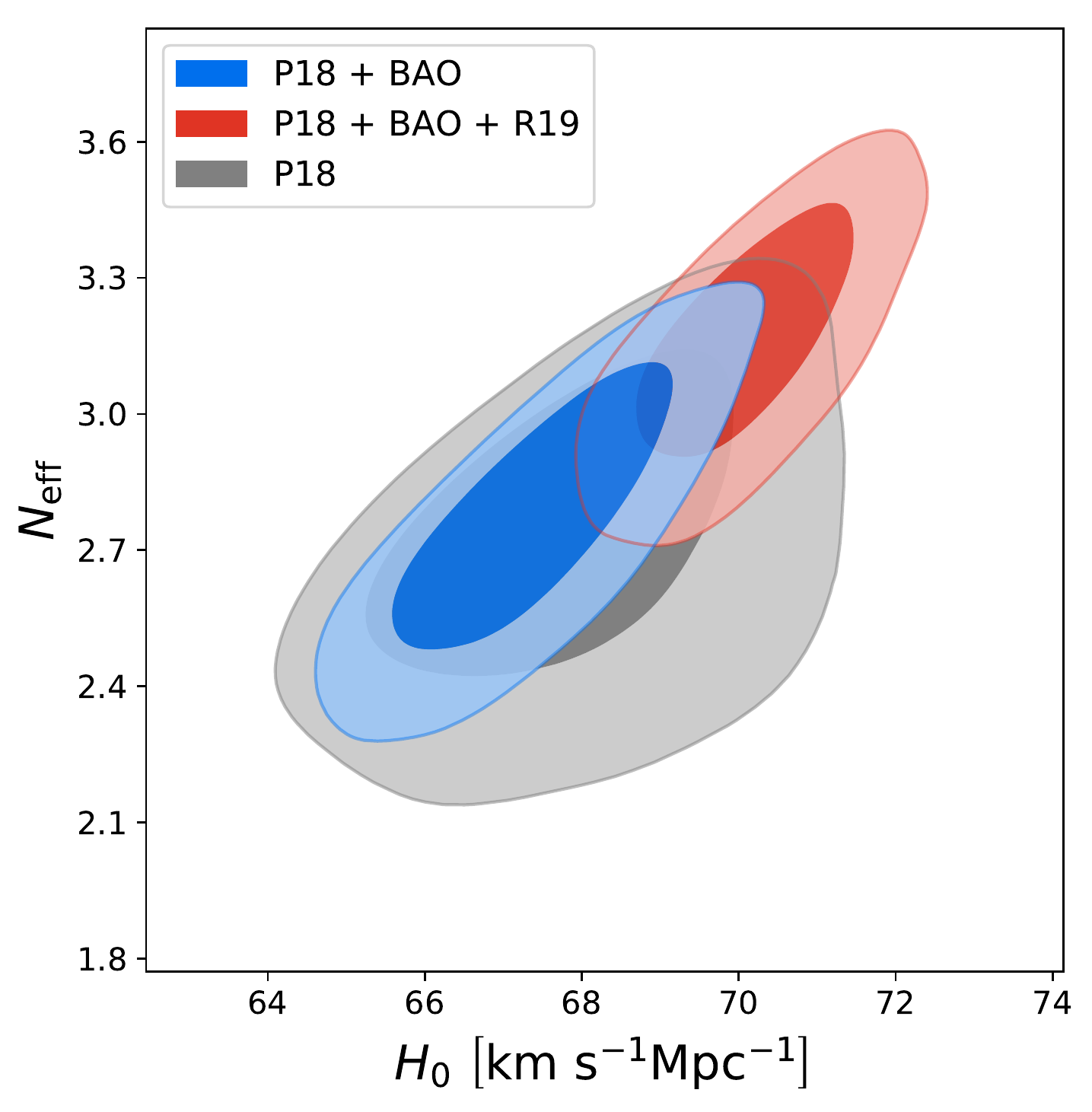}
\includegraphics[width=0.32\textwidth]{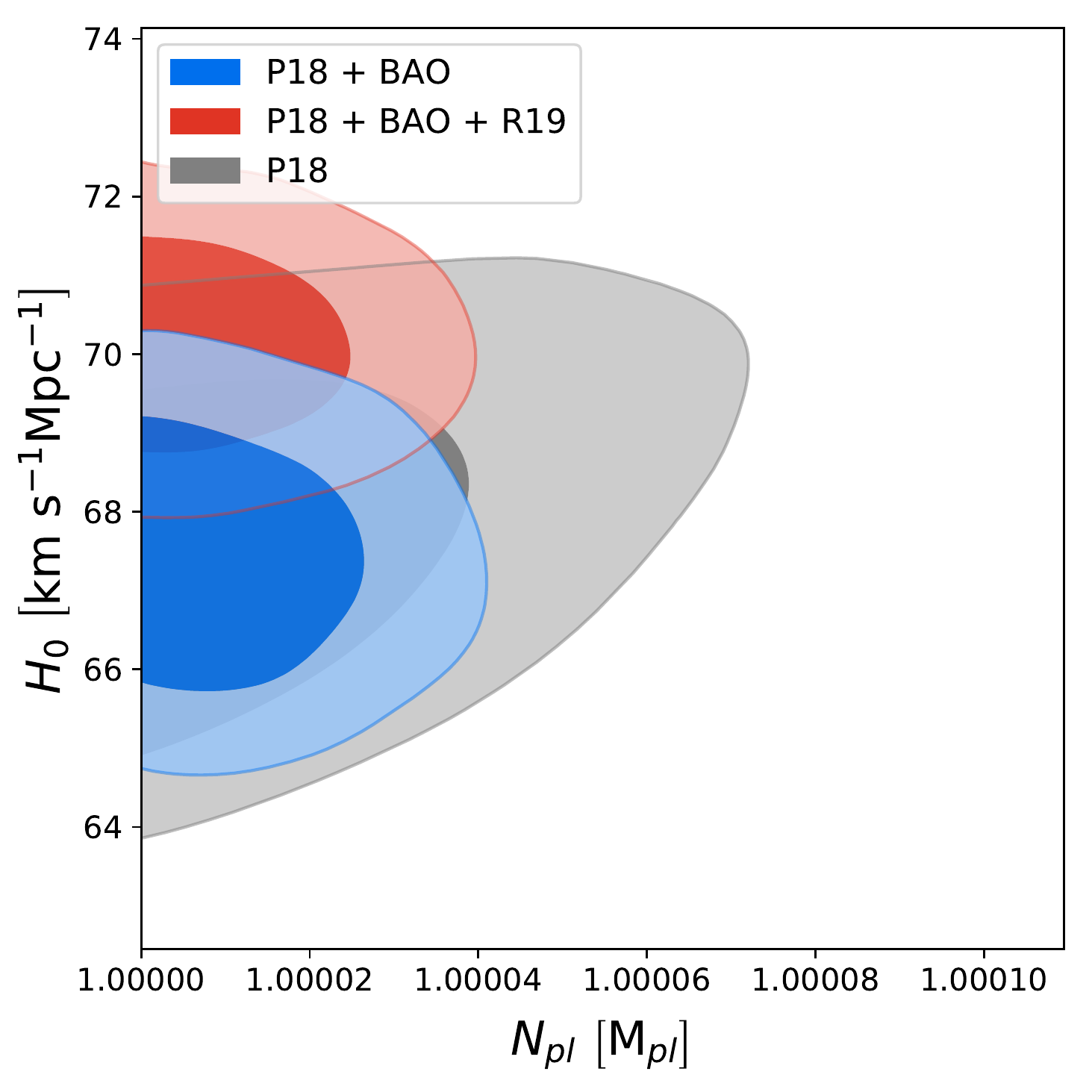}
\caption{Marginalized joint 68\% and 95\% CL regions 2D parameter space using 
the $Planck$ legacy data (gray) in combination with DR12 (blue) and DR12 + R19 (red) 
for the CC+$N_{\rm eff}$ model.}
\label{fig:cc_Neff}
\end{figure}

In the CC model, an analogous correlation in the $N_{\rm eff}-N_{pl}$ parameter space is found, 
see Fig.~\ref{fig:cc_Neff}.
The constraints on $N_{pl}$ are larger, from $N_{pl} < 1.000028$ M$_{pl}$ to 
$N_{pl} < 1.000057$ M$_{pl}$ at 95\% CL with P18 alone and from 
$N_{pl} < 1.000018$ M$_{pl}$ to $N_{pl} < 1.000019$ M$_{pl}$ at 95\% CL 
once we include BAO, see Tab.~\ref{tab:cc_Neff}.

While for the combination P18 + BAO we find a higher value for the Hubble parameter for IG, 
i.e. $H_0 = \left(68.78^{+0.53}_{-0.78}\right)$ km s$^{-1}$Mpc$^{-1}$, and for CC, i.e. 
$H_0 = \left(68.62^{+0.47}_{-0.66}\right)$ km s$^{-1}$Mpc$^{-1}$, compared to the 
$\Lambda$CDM+$N_{\rm eff}$ case, i.e. $H_0 = \left(67.3 \pm 1.1 \right)$ km s$^{-1}$Mpc$^{-1}$, 
the addition of R19 data leads to a closer posterior distribution for $H_0$ 
among the three cases, i.e. $(70.1 \pm 0.8)$ km s$^{-1}$Mpc$^{-1}$ for IG, $(69.6 \pm 0.7)$ 
km s$^{-1}$Mpc$^{-1}$ for CC, and $H_0 = \left(70.0 \pm 0.9 \right)$ km s$^{-1}$Mpc$^{-1}$ 
for $\Lambda$CDM+$N_{\rm eff}$. 
We find a similar posterior distribution also for IG+$N_{\rm eff}$ (CC+$N_{\rm eff}$), 
i.e. $(70.3 \pm 0.9)$ km s$^{-1}$Mpc$^{-1}$ 
($(70.1 \pm 0.9)$ km s$^{-1}$Mpc$^{-1}$), see Fig.~\ref{fig:ig_Neff}.

The addition of BAO data reduces the degeneracy $H_0-\xi\ (-N_{pl})$ increasing the one between 
$N_{\rm eff}-\xi\ (-N_{pl})$ and $H_0-N_{\rm eff}$. In order to reduce comoving sound horizon 
to accommodate a larger value of $H_0$, in this case $N_{\rm eff}$ is moved towards larger 
values, i.e. $3.11 \pm 0.19$ for IG and $3.16 \pm 0.19$ for CC, see Tab.~\ref{tab:ig_Neff}.

\subsection{Neutrino mass} \label{sec:mnu}

The changes in the background evolution caused by neutrino mass, under standard assumptions 
and for a fixed set of standard cosmological parameters, are confined to late times. 
In particular, the neutrino mass impact the angular diameter distance and $z_\Lambda$ 
(the redshift of matter-to-cosmological-constant equality) (see 
Refs.~\cite{Bashinsky:2003tk,Hannestad:2005gj,Hannestad:2006zg,Lesgourgues:2006nd,Wong:2011ip,Lesgourgues:2018ncw,Archidiacono:2016lnv} 
for a review on neutrino mass in cosmology). 

In the standard $\Lambda$CDM scenario, a larger value of $m_\nu$ results in a lower Hubble 
rate inferred from the CMB, exacerbating the $H_0$ {\em tension}. 
However, there is partial correlation between the equation of state of dark energy (DE) $w$ 
and the total neutrino mass $m_\nu$, as first noticed by \cite{Hannestad:2005gj}. 
When $m_\nu$ is increased (or more generally $\Omega_\nu$), $\Omega_m$ can be kept unchanged, 
by simultaneously decreasing $w$, in order to keep the angular diameter 
distance at decoupling fixed. In this case, the impact of neutrino mass on the background is 
confined to variations of $z_\Lambda$ and of the late-time ISW effect.

Cosmological bounds on the neutrino masses can therefore be relaxed by using a DE component 
rather than a cosmological constant. Vice versa, cosmological constraints on the DE 
parameters become larger in comparison to cosmologies with massless neutrinos or with the 
standard minimal assumption of $m_\nu = 0.06$ eV.
The same conclusions have been obtained in the context of Galileon gravity \cite{Barreira:2014ija}.

\begin{figure}
\centering
\includegraphics[width=1.\textwidth]{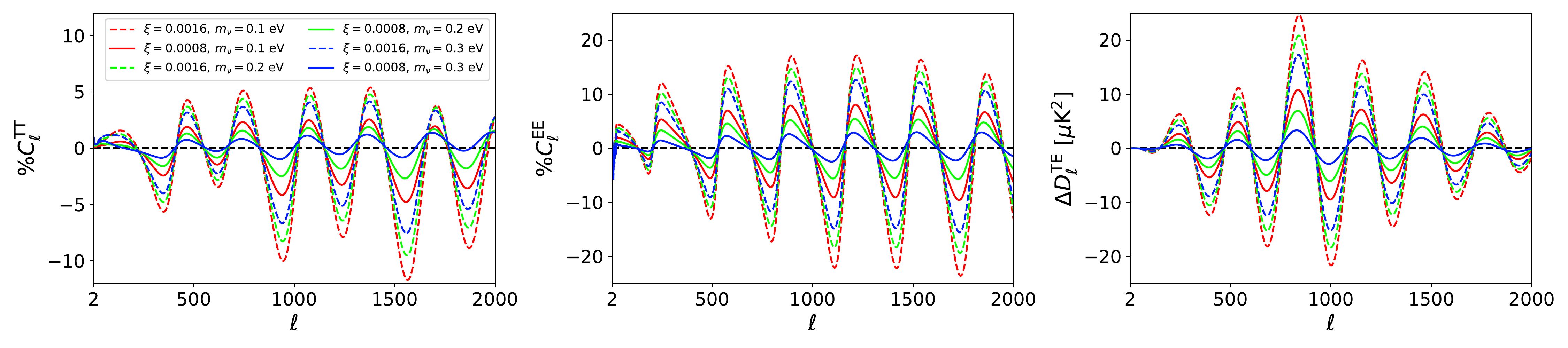}
\includegraphics[width=1.\textwidth]{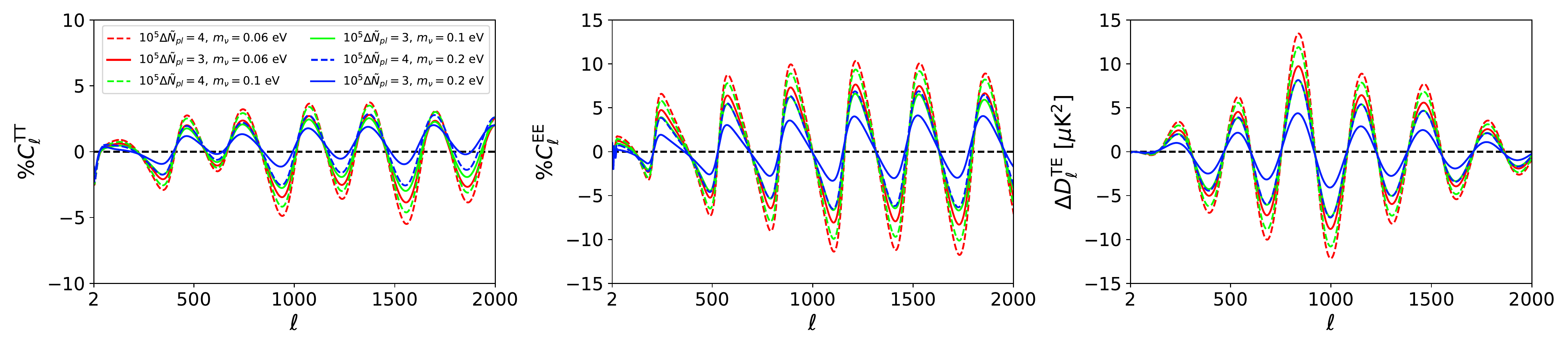}
\caption{Differences with respect to the $\Lambda$CDM with $m_\nu = 0$ eV 
with IG (top panels) for $\xi = 0.0008,\,0.0016$ (solid, dashed) and 
$m_\nu = 0.1,\,0.2,\,0.3$ eV (red, green, blue), and CC (bottom panels) 
for $N_{pl} = 1.00003,\,1.00004$ M$_{pl}$ (solid, dashed) 
and $m_\nu = 0.06,\,0.1,\,0.2$ eV (red, green, blue).}
\label{fig:cl_mnu}
\end{figure}

\begin{figure}
\centering
\includegraphics[width=1.\textwidth]{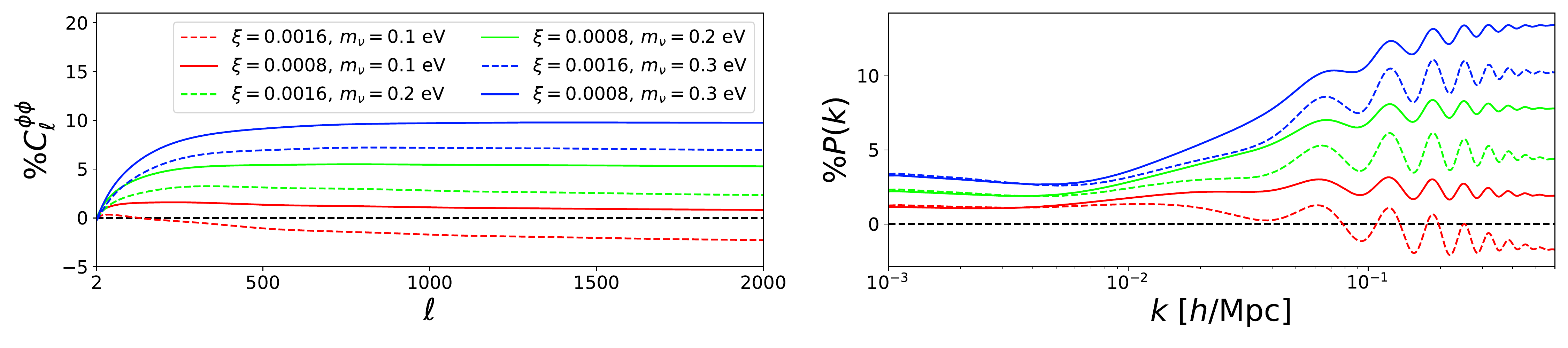}
\includegraphics[width=1.\textwidth]{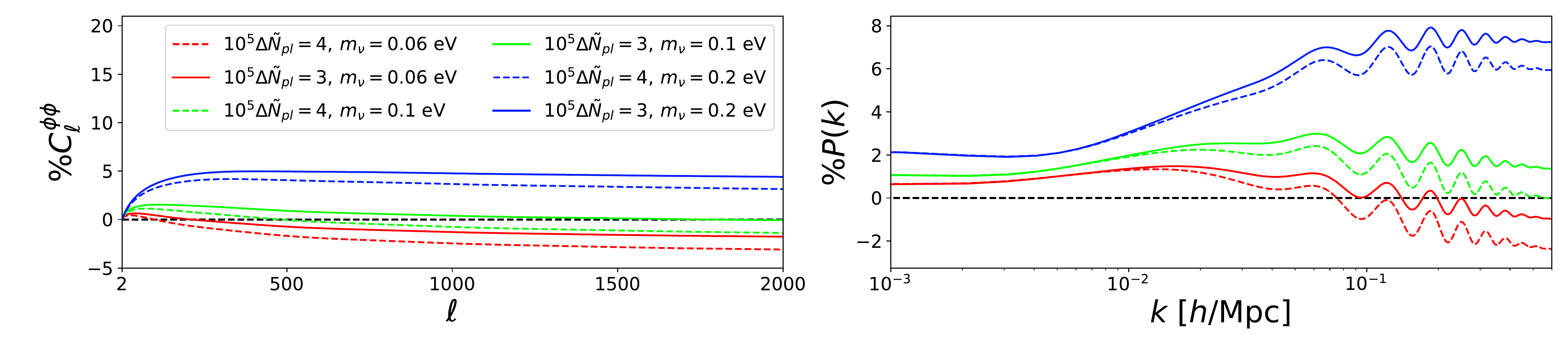}
\caption{Differences with respect to the $\Lambda$CDM with $m_\nu = 0$ eV 
with IG (top panels) for $\xi = 0.0008,\,0.0016$ (solid, dashed) and 
$m_\nu = 0.1,\,0.2,\,0.3$ eV (red, green, blue), and CC (bottom panels) 
for $N_{pl} = 1.00003,\,1.00004$ M$_{pl}$ (solid, dashed) 
and $m_\nu = 0.06,\,0.1,\,0.2$ eV (red, green, blue).}
\label{fig:mpk_mnu}
\end{figure}

We show in Fig.~\ref{fig:cl_mnu} the combined effect on the CMB anisotropies of varying both 
$\xi$ and $m_\nu$ in the IG model. For a fixed value of the coupling parameter $\xi = 0.0008$, 
the differences with respect to the $\Lambda$CDM concordance model are reduced by increasing 
the value of the neutrino mass $m_\nu$ from 0.1 eV to 0.3 eV.
On the late-time observables, i.e. the weak lensing CMB anisotropies and the linear matter power 
spectrum, the partial degeneracy between modified gravity and the neutrino mass is still 
present but with differences concentrated on small scales, see Fig.~\ref{fig:mpk_mnu}.

\begin{figure}
\centering
\includegraphics[height=0.32\textwidth]{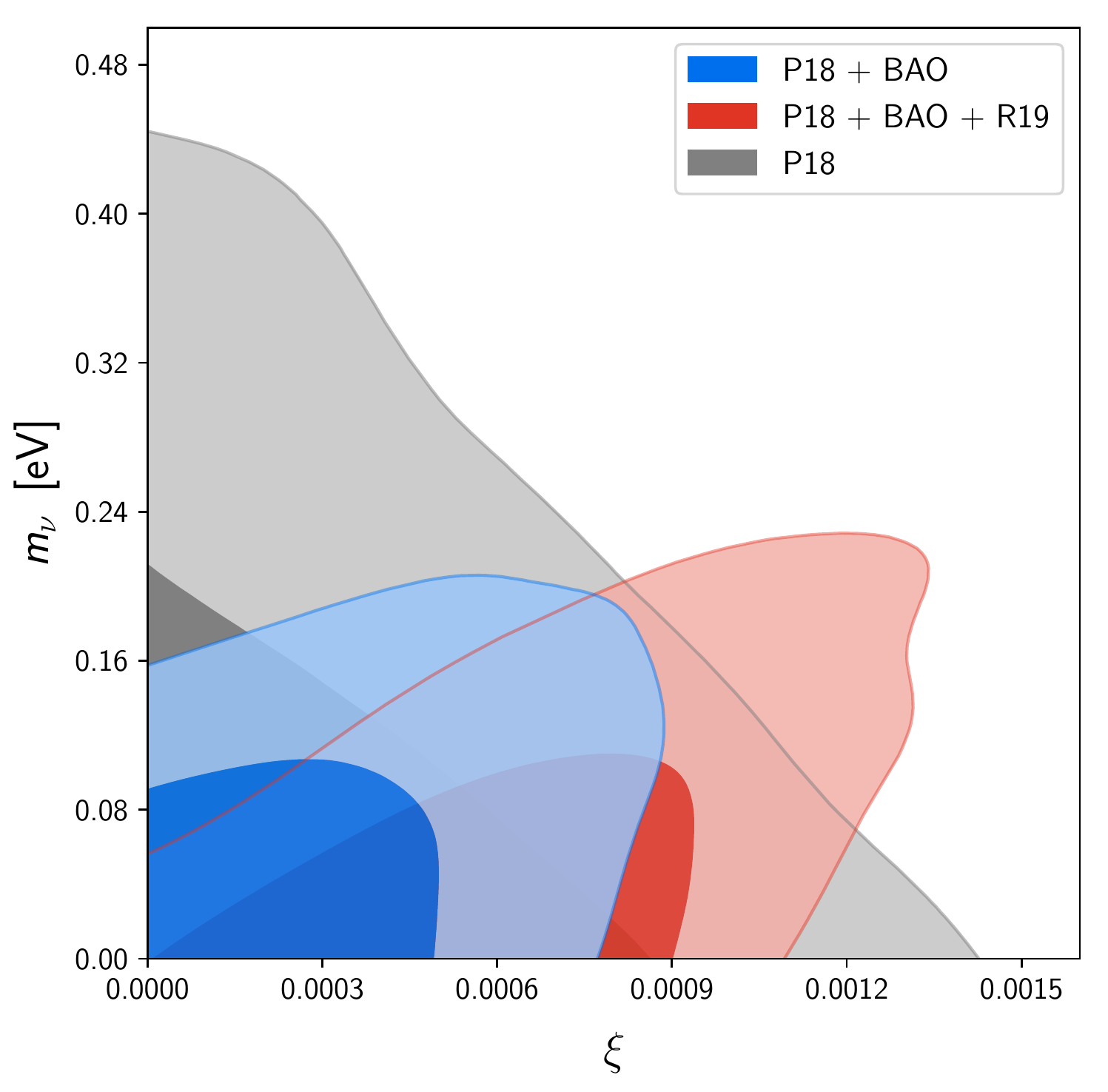}
\includegraphics[height=0.32\textwidth]{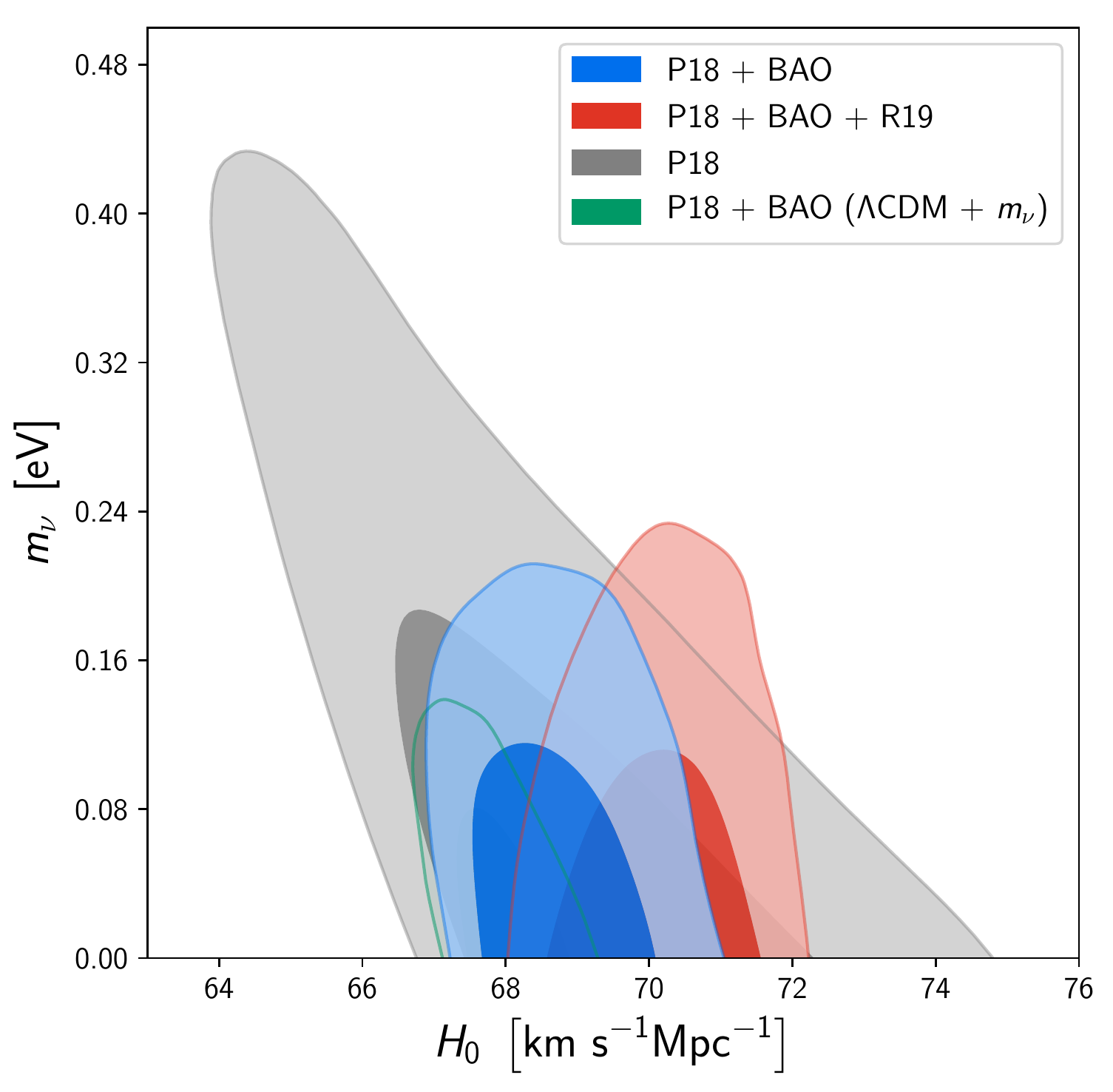}
\includegraphics[height=0.32\textwidth]{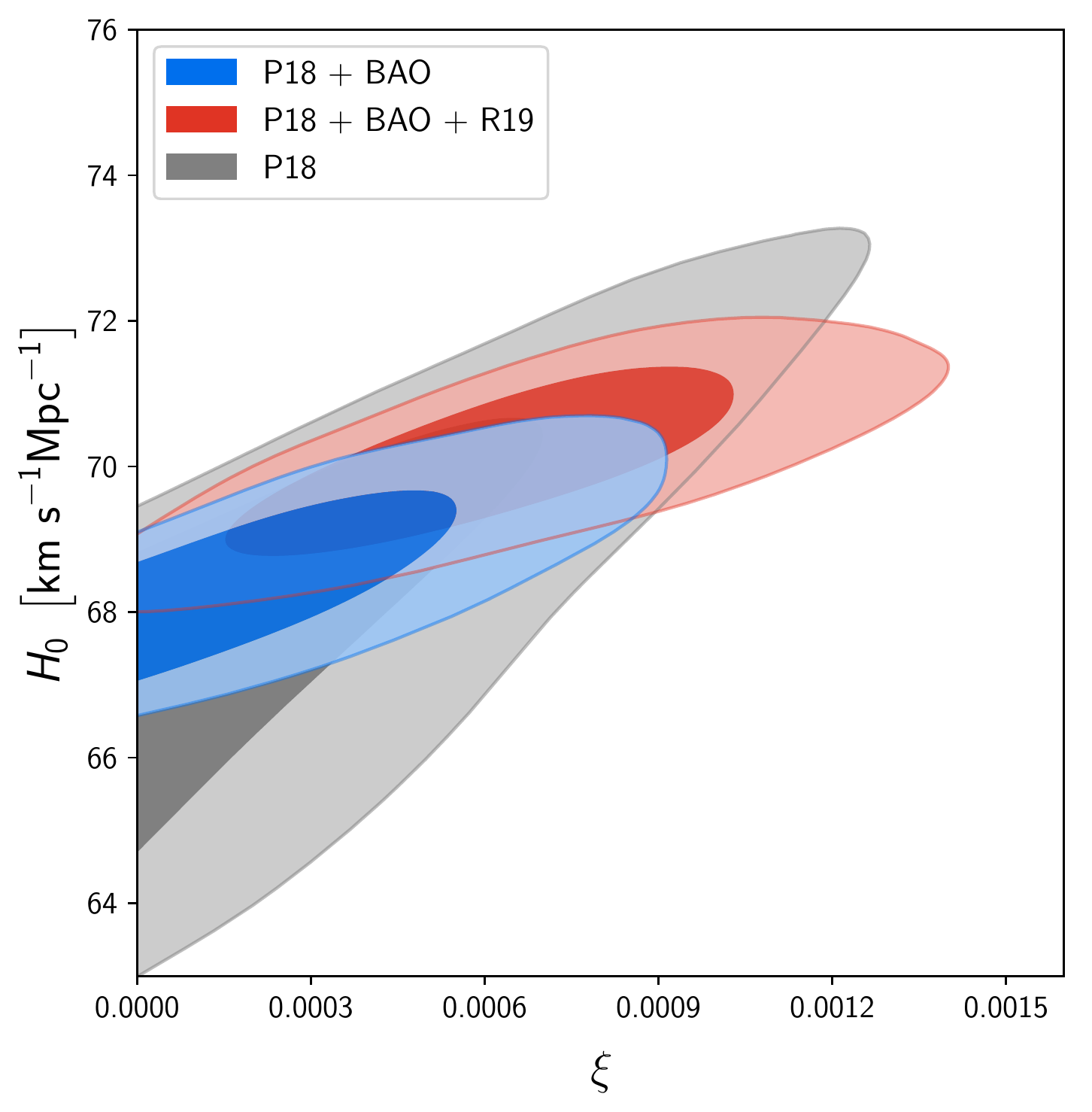}
\caption{Marginalized joint 68\% and 95\% CL regions 2D parameter space using P18 (gray) 
in combination with BAO (blue) and BAO +  R19 (red) for the IG+$m_\nu$ model.
In the central panel, we include the $H_0-N_{\rm eff}$ contours for the $\Lambda$CDM in green.}
\label{fig:ig_mnu}
\end{figure}

In this case the constraints on the coupling parameter $\xi$ become tighter compared to the 
case with $m_\nu = 0$, i.e. from $\xi < 0.00098$ to $\xi < 0.00094$ at 95\% CL for P18\footnote{We 
assume one massive and two massless neutrinos.}.
The CMB anisotropies data prefer to relax the upper bound on the neutrino mass which becomes 
$m_\nu < 0.31$ eV at 95\% CL for P18 29\% larger to the $\Lambda$CDM case $m_\nu < 0.24$ eV.
Including the BAO data, the total neutrino mass is constrained to $m_\nu < 0.17$ eV at 95\% CL, 
42\% larger to the $\Lambda$CDM case $m_\nu < 0.12$ eV, and 
we find $\xi < 0.00076$ at 95\% CL, see Tab.~\ref{tab:ig_mnu}.
The addition of R19 data leads to $H_0 = (70.1 \pm 0.8)$ km s$^{-1}$Mpc$^{-1}$ with 
an upper bound on the total neutrino mass $m_\nu < 0.19$ eV at 95\% CL, 2.5 times larger 
than the limit based on the $\Lambda$CDM model, with a $2\sigma$ detection of the 
coupling parameter $\xi = 0.00065 \pm 0.00057$ at 95\% CL, see Fig.~\ref{fig:ig_mnu}.

\begin{figure}
\centering
\includegraphics[height=0.32\textwidth]{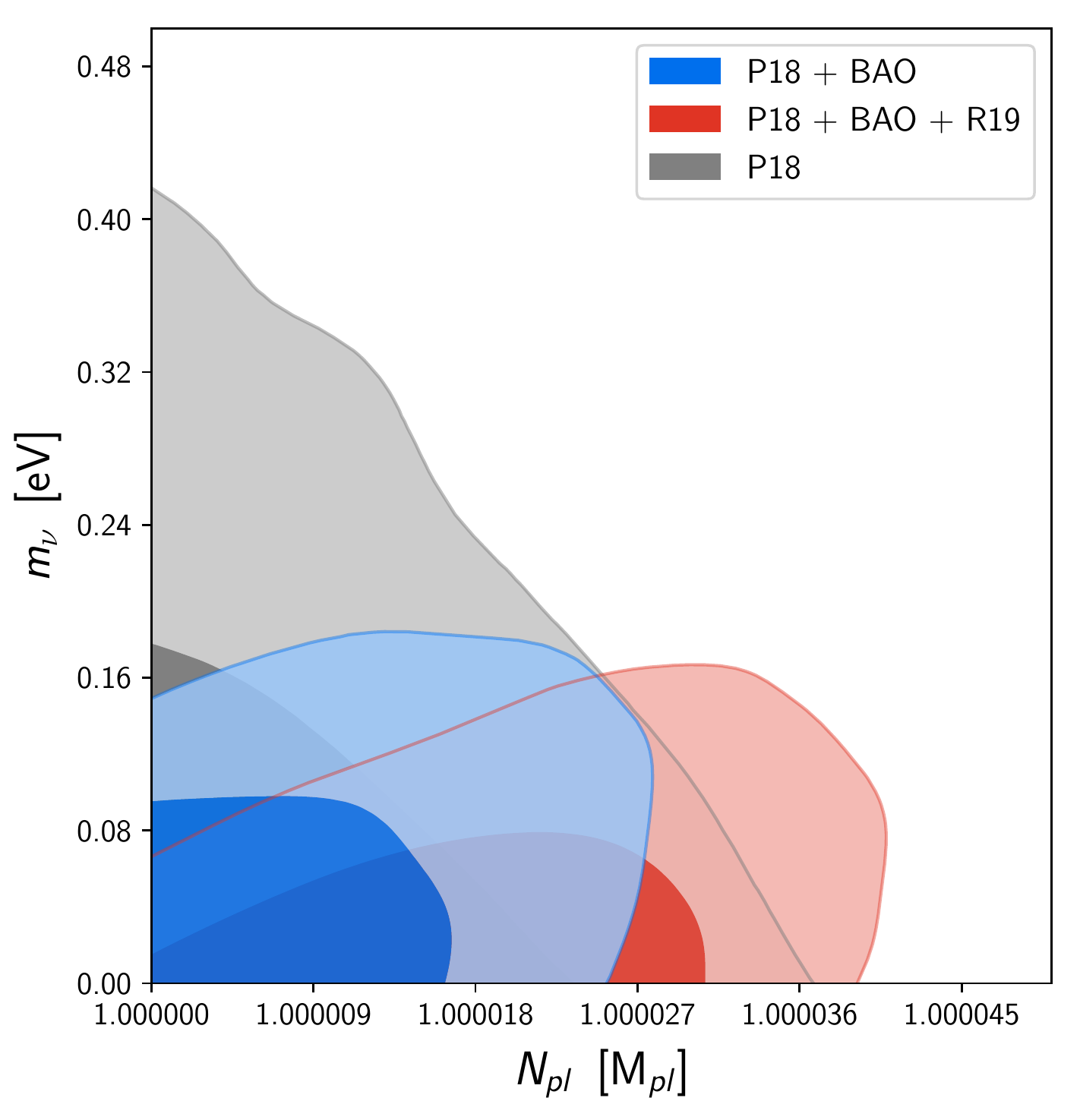}
\includegraphics[height=0.32\textwidth]{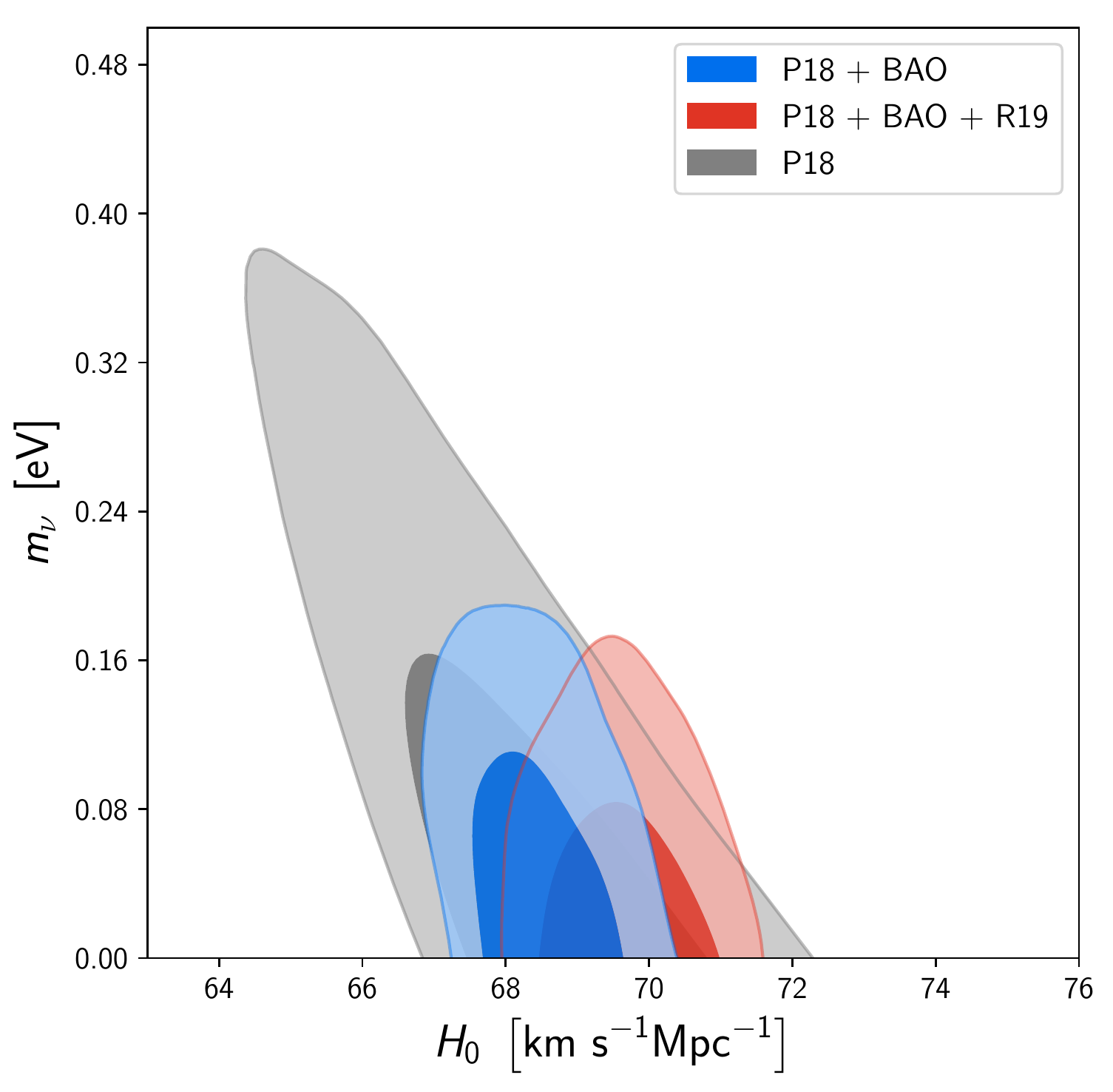}
\includegraphics[height=0.32\textwidth]{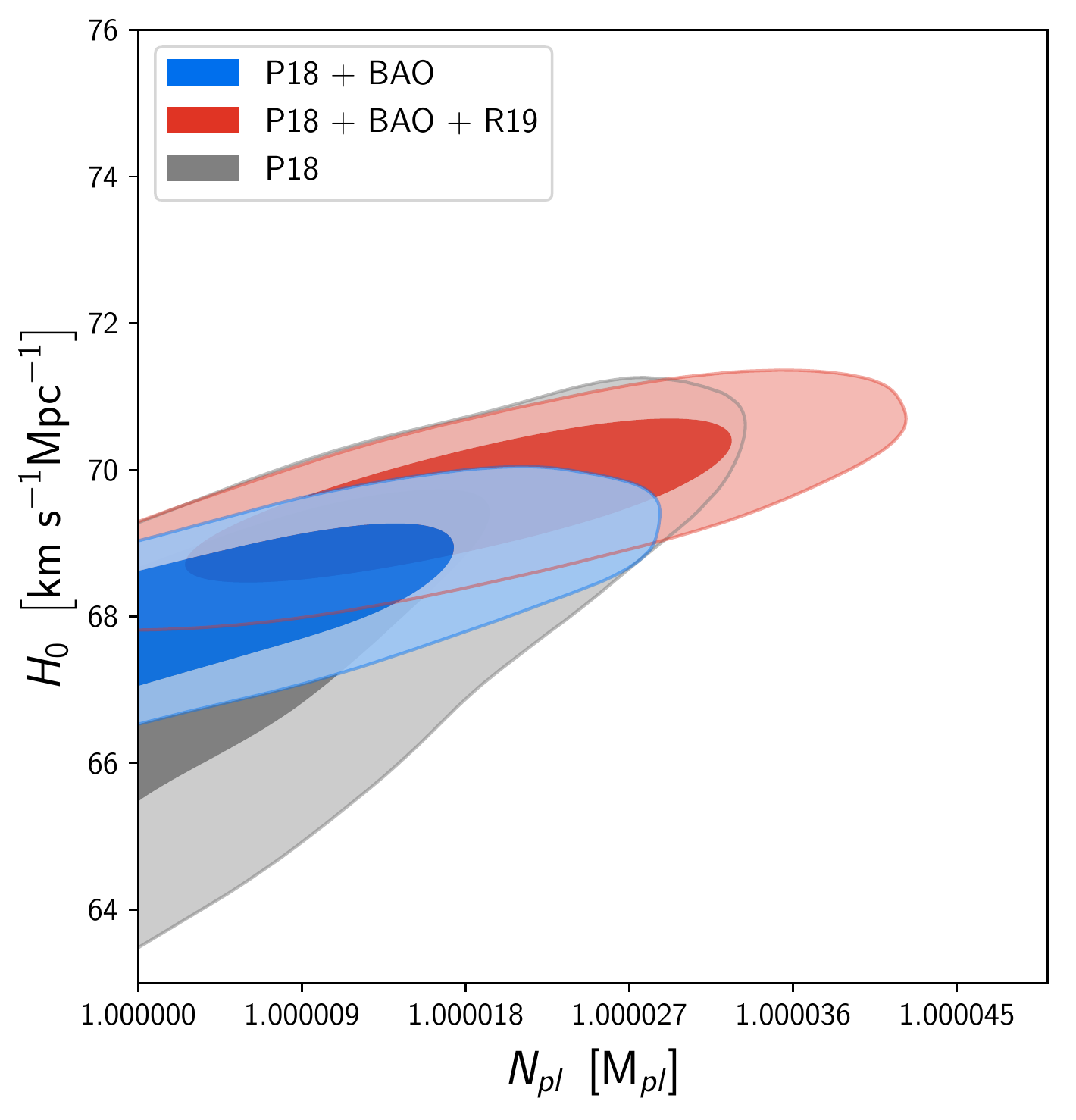}
\caption{Marginalized joint 68\% and 95\% CL regions 2D parameter space using P18 (gray) 
in combination with BAO (blue) and BAO +  R19 (red) for the CC+$m_\nu$ model.}
\label{fig:cc_mnu}
\end{figure}

Analogously, for CC the constraint on $N_{pl}$ becomes tighter compared to the case with 
$m_\nu = 0$, i.e. $N_{pl} < 1.000026\ $M$_{pl}$ for P18 and $N_{pl} < 1.000024\ $M$_{pl}$ 
for P18 + BAO at 95\% CL, see Fig.~\ref{fig:cc_mnu}. Also, for this model, the upper 
bound on the neutrino mass becomes 30\% larger compared to the $\Lambda$CDM case, see 
Tab.~\ref{tab:cc_mnu}.

\subsection{Joint constraints on $N_{\rm eff}$ and neutrino mass} \label{sec:joint}

\begin{figure}
\centering
\includegraphics[width=0.45\textwidth]{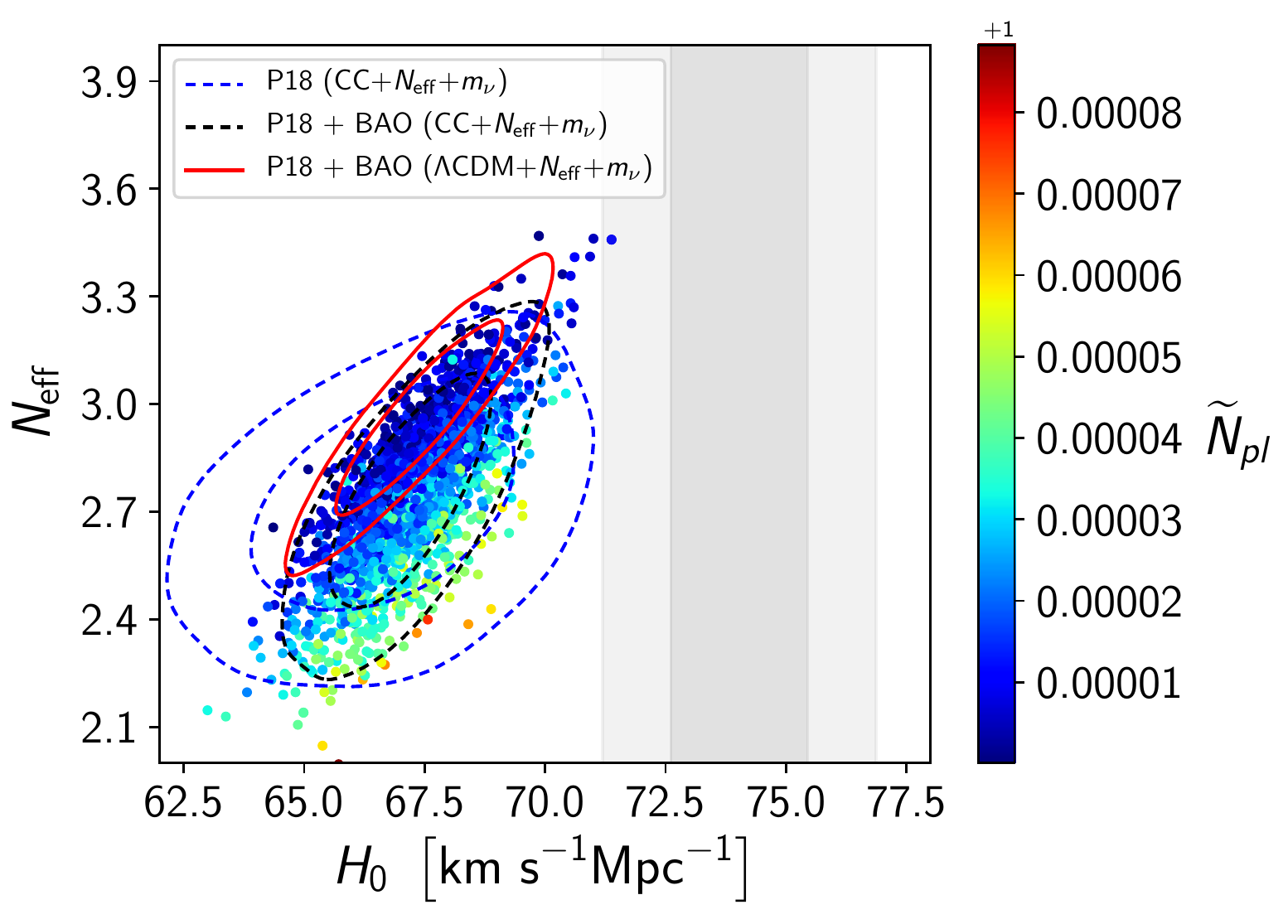}
\includegraphics[width=0.45\textwidth]{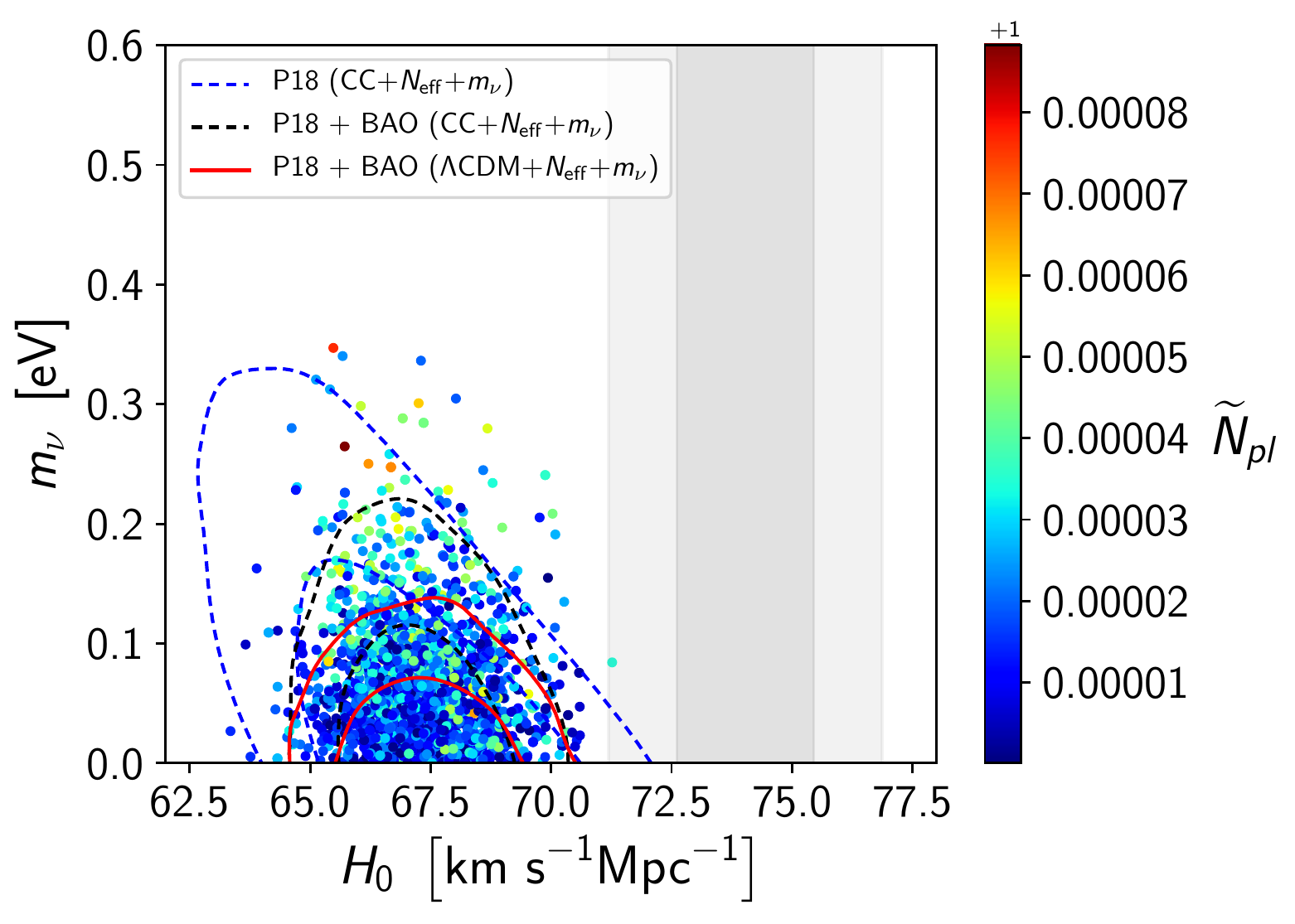}
\caption{Samples of the P18 + BAO chains in the $H_0-N_{\rm eff}$ ($H_0-m_\nu$) plane, 
colour-coded by $N_{pl}$ for the CC+$N_{\rm eff}$+$m_\nu$ model. Dashed blue contours 
show the constraints for CC+$N_{\rm eff}$+$m_\nu$ with P18 alone. Solid red contours 
show the constraints for the $\Lambda$CDM+$N_{\rm eff}$+$m_\nu$ model. The gray bands 
denote the local Hubble parameter measurement from R19 \cite{Riess:2019cxk}.}
\label{fig:cc_tri}
\end{figure}

Finally, we consider the case where both $N_{\rm eff}$ and $m_\nu$ are allowed to vary. 
Despite the larger parameter space and the larger limits on the parameters, 
the models do not accommodate higher values of the Hubble parameter compared 
to the 7- and 8-parameters case analysed before, see Figs.~\ref{fig:ig_tri}-\ref{fig:cc_tri}. 
Moreover, the total neutrino mass is almost uncorrelated with the Hubble parameter.
In this case, the modified gravity 
parameters $\xi$ and $N_{pl}$ are always compatible at $2\sigma$ with the GR limit 
due to the larger parameter space and are given by (see Tabs.~\ref{tab:ig_Neff_mnu}-\ref{tab:cc_Neff_mnu}):

\begin{figure}
\centering
\includegraphics[width=0.45\textwidth]{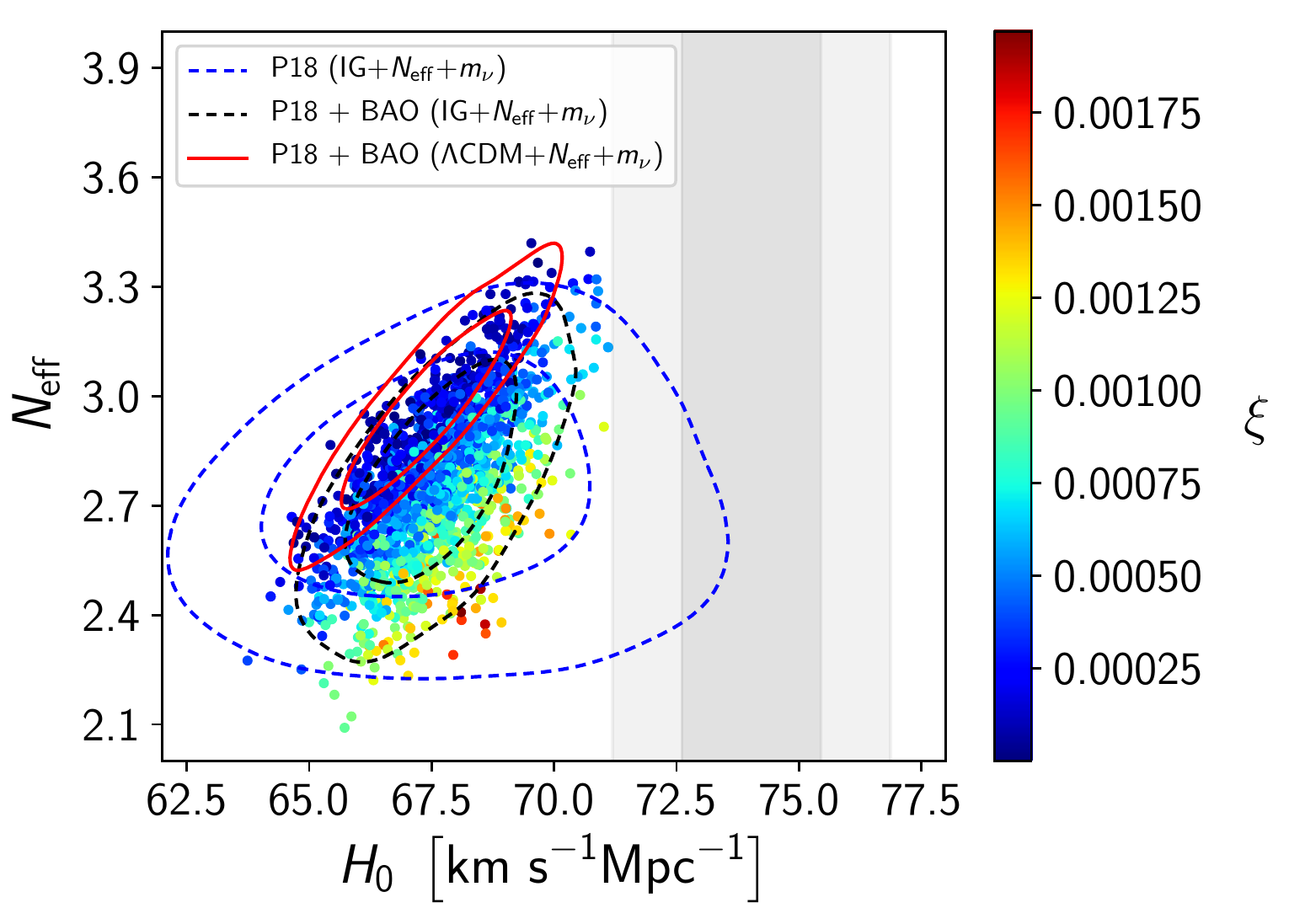}
\includegraphics[width=0.45\textwidth]{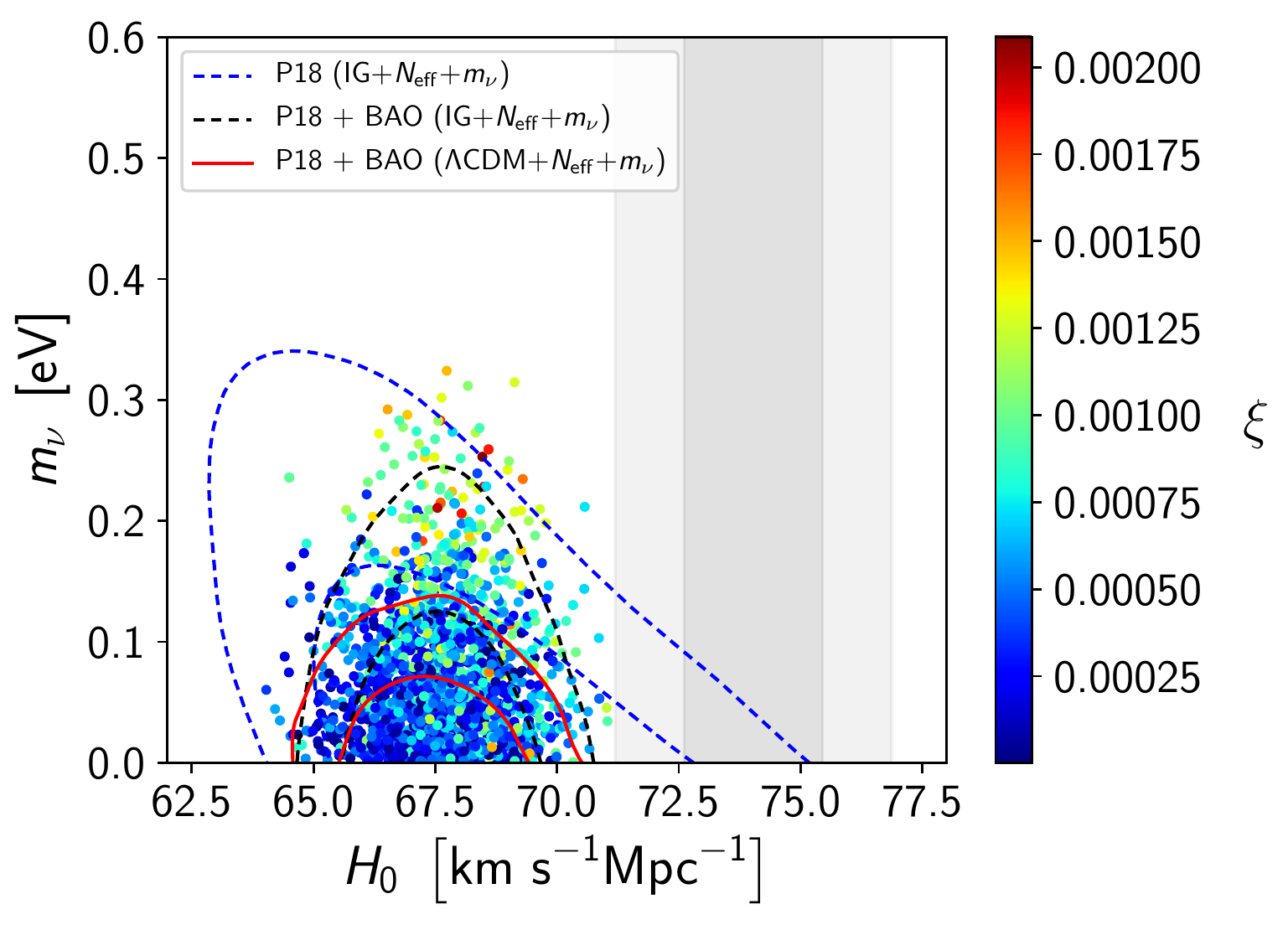}
\caption{Samples of the P18 + BAO chains in the $H_0-N_{\rm eff}$ ($H_0-m_\nu$) plane, 
colour-coded by $\xi$ for the IG+$N_{\rm eff}$+$m_\nu$ model. Dashed blue contours 
show the constraints for IG+$N_{\rm eff}$+$m_\nu$ with P18 alone. Solid red contours 
show the constraints for the $\Lambda$CDM+$N_{\rm eff}$+$m_\nu$ model. The gray bands denote 
the local Hubble parameter measurement from R19 \cite{Riess:2019cxk}.}
\label{fig:ig_tri}
\end{figure}

\begin{equation*}
    \xi < 0.0018\ (95\%\ \text{CL})\,,\qquad N_{\rm eff} = 2.74 \pm 0.22\,,\qquad m_\nu < 0.26\ \text{eV}\ (95\%\ \text{CL})
\end{equation*}
for IG and
\begin{equation*}
    N_{pl} < 1.000050\ \text{M}_{pl}\ (95\%\ \text{CL})\,,\qquad N_{\rm eff} = 2.73 \pm 0.21\,,\qquad m_\nu < 0.26\ \text{eV}\ (95\%\ \text{CL})
\end{equation*}
for the CC case.
When we include BAO data, we obtain 
\begin{equation*}
    \xi < 0.0012\ (95\%\ \text{CL})\,,\qquad N_{\rm eff} = 2.77 \pm 0.20\,,\qquad m_\nu < 0.19\ \text{eV}\ (95\%\ \text{CL})
\end{equation*}
for IG and
\begin{equation*}
    N_{pl} < 1.000042\ \text{M}_{pl}\ (95\%\ \text{CL})\,,\qquad N_{\rm eff} = 2.75 \pm 0.21\,,\qquad m_\nu < 0.17\ \text{eV}\ (95\%\ \text{CL}) 
\end{equation*}
for the CC case.
Adding also R19, we obtain 
\begin{equation*}
    \xi < 0.0013\ (95\%\ \text{CL})\,,\qquad N_{\rm eff} = 3.08 \pm 0.20\,,\qquad m_\nu < 0.19\ \text{eV}\ (95\%\ \text{CL})
\end{equation*}
for IG and
\begin{equation*}
    N_{pl} < 1.000040\ \text{M}_{pl}\ (95\%\ \text{CL})\,,\qquad N_{\rm eff} = 3.14 \pm 0.20\,,\qquad m_\nu < 0.14\ \text{eV}\ (95\%\ \text{CL}) 
\end{equation*}
for the CC case.

\section{Conclusions} \label{sec:conclusion}

We have studied the simplest class of scalar-tensor theories of gravity where Newton’s 
constant is allowed to vary in time and space. 
These non-minimally coupled theories are characterized by a coupling to the Ricci scalar 
$R$ of the type $F(\sigma) = N^2_{pl} + \xi \sigma^2$, which contain the {\em induced gravity} 
\cite{Cooper:1982du,Wetterich:1987fk,Finelli:2007wb} model for $N_{pl}=0$ and $\xi > 0$, 
and the {\em conformal coupling} model \cite{Rossi:2019lgt} for $\xi = -1/6$.
These models contribute to the radiation density budget 
in the radiation-dominated epoch and change the background expansion history compared 
to the $\Lambda$CDM concordance model naturally alleviating the $H_0$ {\em tension}. 

Compared to previous constraints \cite{Ballardini:2016cvy,Rossi:2019lgt}, the improvement of 
{\em Planck} 2018 polarization data lead to tighter results, i.e. 
$\xi < 0.00098$ and $N_{pl} < 1.000028$ M$_{pl}$ both at 95\% CL. When BAO data from BOSS DR12 
are added, we obtain tighter limits, i.e. $\xi < 0.00055$ and $N_{pl} < 1.000018$ M$_{pl}$. 
For $\xi=-1/6$, P18 and BAO data lead to constraints on the post-Newtonian parameters which are 
tighter than those derived within the Solar System.

It is interesting to note that for CMB and BAO data these models allow for 
values of $H_0$ larger than $\Lambda$CDM. 
Using P18 data, we find $H_0 = \left(69.6^{+0.8}_{-1.7}\right)$ km s$^{-1}$Mpc$^{-1}$ 
($H_0 = \left(69.0^{+0.7}_{-1.2}\right)$ km s$^{-1}$Mpc$^{-1}$) for the IG (CC) case 
compared to $H_0 = (67.36 \pm 0.54)$ km s$^{-1}$Mpc$^{-1}$ for $\Lambda$CDM, alleviating 
the tension from $4.4\sigma$ to 
2.7$\sigma$ (3.2$\sigma$) for P18 and 3.5$\sigma$ (3.6$\sigma$) including BAO for 
IG (CC).

Including BAO and R19, we obtain $H_0 = \left(70.06 \pm 0.81\right)$ km s$^{-1}$Mpc$^{-1}$ 
($H_0 = \left(69.64^{+0.65}_{-0.73}\right)$ km s$^{-1}$Mpc$^{-1}$) for IG (for CC). 
The value for $H_0$ we find is similar to what can be obtained in other models which aim to 
solve the $H_0$ tension, but the models considered here have just one extra parameter as 
$\Lambda$CDM+$N_{\rm eff}$. 
Similar valus of $H_0$ can also be found for NMC beyond the IG and CC cases considered here 
\cite{Braglia:2020iik}.

We have extended our analysis to a general neutrino sector by allowing the effective number 
of relativistic species $N_{\rm eff}$ and the neutrino mass $m_\nu$ to vary.
Both $N_{\rm eff}$ and $m_\nu$ are partially degenerate with the deviations from GR, as happens in 
other modified gravity models \cite{Motohashi:2012wc,Barreira:2014ija,Chudaykin:2014oia}. 
Whereas $N_{\rm eff}$ and the scalar field act as an additional source of radiation in the early 
Universe, at late times the background contribution to $\Omega_m$ due to $m_\nu$ can be  
compensated from the scalar field in order to keep the angular diameter distance at decoupling fixed, 
see Figs.~\ref{fig:cl_Neff}-\ref{fig:cl_mnu}-\ref{fig:mpk_mnu}. We have shown however that these are 
only partial degeneracies which could be broken by combination of observations at different redshifts. 

In case with $N_{\rm eff}$ (Sec.~\ref{sec:Neff}) the limit on $\xi$ becomes $\sim 94\%$ ($\sim 42\%$) 
larger with P18 (P18+BAO) while the mean on the number of neutrinos moves around $1\sigma$ towards 
lower values compared to the $\Lambda$CDM case without significantly degrading its uncertainty, 
i.e. $N_{\rm eff} = 2.79 \pm 0.20$ ($N_{\rm eff} = 2.85 \pm 0.17$).
For CC the limit on $N_{pl}$ becomes $\sim 104\%$ ($\sim 6\%$) larger with P18 (P18+BAO) 
and analogously to IG we find $N_{\rm eff} = 2.73^{+0.25}_{-0.22}$ ($N_{\rm eff} = 2.81 \pm 0.19$).

The upper bound on the neutrino mass (Sec.~\ref{sec:mnu}) is $\sim 29\%$ ($\sim 42\%$) is also degraded 
with P18 (P18+BAO) compared to the $\Lambda$CDM case, i.e. $m_\nu < 0.31$ eV ($m_\nu < 0.17$ eV), 
whereas the constraint on $\xi$ is slightly tighter with CMB data alone in order to relax the 
constraint on $m_\nu$.
Analogously, for CC the limit on the neutrino mass is $\sim 17\%$ ($\sim 33\%$) larger with P18 
(P18+BAO) compared to the $\Lambda$CDM case.
When both $N_{\rm eff}$ and $m_\nu$ are allowed to vary, we see that the constraints on $\xi$ 
and $N_{\rm pl}$ degrade by a factor two compared to the case with $N_{\rm eff} = 3.046$ and $m_\nu = 0$ eV 
also in presence of BAO data, i.e. $\xi < 0.0012$ and $N_{pl} < 1.000042$ M$_{pl}$ at 95\% CL.
For the data used, the combination of the modification to gravity in our models to non-standard 
neutrino physics does not lead to higher values of $H_0$ compared to the case with standard 
assumptions in the neutrino sector.

\section*{Acknowledgments}
We wish to thank Maria Archidiacono and Thejs Brinckmann for useful suggestions on the use of 
{\tt MontePython}.
MBa, MBr, FF, DP acknowledge financial contribution from the contract ASI/INAF for the Euclid 
mission n.2018-23-HH.0.
FF and DP acknowledge financial support by ASI Grant 2016-24-H.0. 
AAS was partially supported by the Russian Foundation for Basic Research grant No. 20-02-00411.

    \newpage

\appendix
\section{Tables} \label{sec:appendix}

\subsection{Updated {\em Planck} 2018 results}

\begin{table*}[h!]
{\small
\centering
\begin{tabular}{l|ccc}
\hline
\hline
                                         & P18 & P18 + BAO & P18 + BAO + R19  \\
\hline
$\omega_{\rm b}$                         & $0.02244_{-0.00016}^{+0.00014}$    & $0.02239 \pm 0.00013$                 & $0.02246 \pm 0.00013$  \\
$\omega_{\rm c}$                         & $0.1198 \pm 0.0012$                & $0.1201 \pm 0.0011$                   & $0.1200 \pm 0.0011$  \\
$H_0$ [km s$^{-1}$Mpc$^{-1}$]            & $69.6^{+0.8}_{-1.7}\ (2.7\sigma)$  & $68.78^{+0.53}_{-0.78}\ (3.5\sigma)$  & $70.06 \pm 0.81\ (2.4\sigma)$ \\
$\tau$                                   & $0.0551^{+0.0065}_{-0.0078}$       & $0.0545^{+0.0063}_{-0.0071}$          & $0.0554^{+0.0064}_{-0.0073}$  \\
$\ln \left(  10^{10} A_{\rm s} \right)$  & $3.047^{+0.014}_{-0.015}$          & $3.046 \pm 0.013$                     & $3.049 \pm 0.013$  \\
$n_{\rm s}$                              & $0.9680^{+0.0044}_{-0.0052}$       & $0.9662 \pm 0.0038$                   & $0.9688 \pm 0.0037$  \\
$\zeta_{\rm IG}$                         & $< 0.0039$ (95\% CL)               & $< 0.0022$ (95\% CL)                  & $0.00202^{+0.00090}_{-0.00100}$  \\
\hline
$\xi$                                                     & $< 0.00098$ (95\% CL)  & $< 0.00055$ (95\% CL)  & $0.00051^{+0.00043}_{-0.00046}$ (95\% CL)  \\
$\gamma_{PN}$                                             & $> 0.9961$ (95\% CL)   & $> 0.9978$ (95\% CL)   & $0.9980^{+0.0010}_{-0.0009}$  \\
$\delta G_\mathrm{N}/G_\mathrm{N}$ (z=0)                  & $> -0.029$ (95\% CL)   & $> -0.016$ (95\% CL)   & $-0.0149 \pm 0.0068$  \\
$10^{13} \dot{G}_\mathrm{N}/G_{\rm N}$ (z=0) [yr$^{-1}$]  & $> -1.16$ (95\% CL)    & $> -0.66$ (95\% CL)    & $-0.61 \pm 0.28$  \\
$G_\mathrm{N}/G$ (z=0)                                    & $> 0.9981$ (95\% CL)   & $> 0.9989$ (95\% CL)   & $0.99899_{-0.00045}^{+0.00050}$  \\
\hline
$\Omega_{\rm m}$                        & $0.2940^{+0.0150}_{-0.0095}$  & $0.3013^{+0.0072}_{-0.0062}$  & $0.2903 \pm 0.0068$  \\
$\sigma_8$                              & $0.8347_{-0.0130}^{+0.0074}$  & $0.8308_{-0.0096}^{+0.0067}$  & $0.840 \pm 0.010$  \\
$r_s$ [Mpc]                             & $146.37_{-0.40}^{+0.79}$      & $146.63_{-0.34}^{+0.55}$      &    $146.03_{-0.59}^{+0.67}$  \\
\hline
$\Delta \chi^2$                         & $0.2$ & $0.2$  & $-3.1$  \\
\hline
\hline
\end{tabular}}
\caption{\label{tab:ig} 
Constraints on main and derived parameters (at 68\% CL if not otherwise stated) considering 
P18 in combination with BAO and BAO + R19 for the IG model.}
\end{table*}

\begin{table*}[h!]
{\small
\centering
\begin{tabular}{l|ccc}
\hline
\hline
                                         & P18 & P18 + BAO & P18 + BAO + R19  \\
\hline
$\omega_{\rm b}$                         & $0.02244 \pm 0.00015$              & $0.02241 \pm 0.00013$                 & $0.0250 \pm 0.0013$  \\
$\omega_{\rm c}$                         & $0.1197 \pm 0.0012$                & $0.11990 \pm 0.00094$                 & $0.1195 \pm 0.0010$  \\
$H_0$ [km s$^{-1}$Mpc$^{-1}$]            & $69.0^{+0.7}_{-1.2}\ (3.2\sigma)$  & $68.62_{-0.66}^{+0.47}\ (3.6\sigma)$  & $69.64^{+0.65}_{-0.73}\ (2.8\sigma)$  \\
$\tau$                                   & $0.0554^{+0.0064}_{-0.0081}$       & $0.0551_{-0.0076}^{+0.0058}$          & $0.0562^{+0.0066}_{-0.0077}$  \\
$\ln \left(  10^{10} A_{\rm s} \right)$  & $3.048^{+0.013}_{-0.016}$          & $3.047_{-0.015}^{+0.011}$             & $3.050^{+0.013}_{-0.015}$  \\
$n_{\rm s}$                              & $0.9684 \pm 0.0047$                & $0.9668 \pm 0.0039$                   & $0.9707 \pm 0.0040$  \\
$N_{\rm pl}$ [M$_{\rm pl}$]              & $< 1.000028$    (95\% CL)          & $< 1.000018$ (95\% CL)                & $< 1.000031$ (95\% CL)  \\
\hline
$\gamma_{PN}$                                             & $> 0.999972$   (95\% CL)          & $> 0.999982$ (95\% CL)            & $> 0.999969$ (95\% CL)  \\
$\beta_{PN}$                                              & $< 1.0000023$  (95\% CL)          & $< 1.0000015$ (95\% CL)           & $< 1.0000025$ (95\% CL)  \\
$\delta G_\mathrm{N}/G_\mathrm{N}$ (z=0)                  & $> -0.026$ (95\% CL)              & $> -0.017$ (95\% CL)              & $> -0.029$ (95\% CL)  \\
$10^{13} \dot{G}_\mathrm{N}/G_{\rm N}$ (z=0) [yr$^{-1}$]  & $> -3.8\times 10^{-9}$ (95\% CL)  & $> -2.5\times 10^{-9}$ (95\% CL)  & $> -4.2\times 10^{-9}$ (95\% CL)  \\
$G_\mathrm{N}/G$ (z=0)                                    & $> 0.999986$ (95\% CL)            & $> 0.999991$ (95\% CL)            & $> 0.999985$ (95\% CL)  \\
\hline
$\Omega_{\rm m}$                        & $0.299_{-0.009}^{+0.011}$  & $0.3023 \pm 0.0061$           & $0.2928 \pm 0.0064$  \\
$\sigma_8$                              & $0.832_{-0.007}^{+0.011}$  & $0.8299_{-0.0088}^{+0.0060}$  & $0.8364_{-0.011}^{+0.0089}$  \\
$r_s$ [Mpc]                             & $146.71_{-0.33}^{+0.46}$   & $146.82^{+0.37}_{-0.28}$      & $146.53^{+0.51}_{-0.42}$  \\
\hline
$\Delta \chi^2$                         & $2.2$ & $0.8$ & $-1.7$  \\
\hline
\hline
\end{tabular}}
\caption{\label{tab:cc} 
Constraints on main and derived parameters (at 68\% CL if not otherwise stated) considering 
P18 in combination with BAO and BAO + R19 for the CC model.}
\end{table*}

    \newpage

\subsection{Degeneracy with the neutrino sector: $N_{\rm eff}$}

\begin{table*}[h!]
{\small
\centering
\begin{tabular}{l|ccc}
\hline
\hline
                                         & P18 & P18 + BAO & P18 + BAO + R19  \\
\hline
$\omega_{\rm b}$                         & $0.02227_{-0.00021}^{+0.00018}$     & $0.02225 \pm 0.00019$              & $0.02250 \pm 0.00019$  \\
$\omega_{\rm c}$                         & $0.1161 \pm 0.0031$                 & $0.1172 \pm 0.0030$                & $0.1210 \pm 0.0029$  \\
$H_0$ [km s$^{-1}$Mpc$^{-1}$]            & $69.2_{-2.4}^{+1.5}\ (2.3\sigma)$   & $67.9_{-1.2}^{+1.0}\ (3.5\sigma)$  & $70.28 \pm 0.92\ (2.2\sigma)$  \\
$\tau$                                   & $0.0547 \pm 0.0078$                 & $0.0526 \pm 0.0069$                & $0.0549 \pm 0.0072$  \\
$\ln \left(  10^{10} A_{\rm s} \right)$  & $3.038 \pm 0.016$                   & $3.035 \pm 0.015$                  & $3.050 \pm 0.016$  \\
$n_{\rm s}$                              & $0.9617_{-0.0088}^{+0.0049}$        & $0.9600^{+0.0045}_{-0.0079}$       & $0.9707 \pm 0.0069$  \\
$\zeta_{\rm IG}$                         & $< 0.0076$ (95\% CL)                & $< 0.0031$ (95\% CL)               & $< 0.0040$ (95\% CL)  \\
$N_{\rm eff}$                            & $2.79 \pm 0.20$                     & $2.85 \pm 0.17$                    & $3.11 \pm 0.19$  \\
\hline
$\xi$                                                     & $< 0.0019$ (95\% CL)  & $< 0.00078$ (95\% CL)  & $< 0.0010$ (95\% CL)  \\
$\gamma_{PN}$                                             & $> 0.9925$ (95\% CL)  & $> 0.9969$ (95\% CL)   & $> 0.9960$ (95\% CL)  \\
$\delta G_\mathrm{N}/G_\mathrm{N}$ (z=0)                  & $> -0.055$ (95\% CL)  & $> -0.023$ (95\% CL)   & $> -0.029$ (95\% CL)  \\
$10^{13} \dot{G}_\mathrm{N}/G_{\rm N}$ (z=0) [yr$^{-1}$]  & $> -2.2$ (95\% CL)    & $> -0.93$ (95\% CL)    & $> -1.2$ (95\% CL)  \\
$\delta G_\mathrm{N}/G$ (z=0)                             & $> 0.9962$ (95\% CL)  & $> 0.9985$ (95\% CL)   & $> -0.9980$ (95\% CL)  \\
\hline
$\Omega_{\rm m}$                           & $0.290_{-0.012}^{+0.022}$  & $0.3022 \pm 0.0074$  & $0.2906 \pm 0.0067$  \\
$\sigma_8$                                 & $0.834_{-0.018}^{+0.012}$  & $0.825 \pm 0.010$    & $0.841 \pm 0.010$  \\
$r_s$ [Mpc]                                & $148.2_{-1.5}^{+1.8}$      & $148.4 \pm 1.7$      & $145.5 \pm 1.5$  \\
\hline
$\Delta \chi^2$                            & $1.7$ & $-1.8$ & $-3.0$  \\
\hline
\hline
\end{tabular}}
\caption{Constraints on main and derived parameters (at 68\% CL if not otherwise stated) considering 
P18 in combination with BAO and BAO + R19 for the IG+$N_{\rm eff}$ model.}
\label{tab:ig_Neff} 
\end{table*}

\begin{table*}[h!]
{\small
\centering
\begin{tabular}{l|ccc}
\hline
\hline
                                         & P18 & P18 + BAO & P18 + BAO + R19  \\
\hline
$\omega_{\rm b}$                         & $0.02223 \pm 0.00022$         & $0.02215 \pm 0.00022$         & $0.02257 \pm 0.00018$  \\
$\omega_{\rm c}$                         & $0.1151 \pm 0.0033$           & $0.1162 \pm 0.0031$           & $0.1213 \pm 0.0030$  \\
$H_0$ [km s$^{-1}$Mpc$^{-1}$]            & $67.9 \pm 1.4\ (3.1\sigma)$   & $67.1 \pm 1.2\ (3.7\sigma)$   & $70.10 \pm 0.92\ (2.0\sigma)$  \\
$\tau$                                   & $0.0539_{-0.0074}^{+0.0060}$  & $0.0544_{-0.0074}^{+0.0061}$  & $0.0561^{+0.0063}_{-0.0075}$  \\
$\ln \left(  10^{10} A_{\rm s} \right)$  & $3.034 \pm 0.017$             & $3.035 \pm 0.016$             & $3.053^{+0.014}_{-0.016}$  \\
$n_{\rm s}$                              & $0.9598 \pm 0.0084$           & $0.9606 \pm 0.0071$           & $0.9736 \pm 0.0062$  \\
$N_{\rm pl}$ [M$_{\rm pl}$]              & $< 1.000057$ (95\% CL)        & $< 1.000019$ (95\% CL)        & $< 1.000032$ (95\% CL)  \\
$N_{\rm eff}$                            & $2.73^{+0.25}_{-0.22}$        & $2.81 \pm 0.19$               & $3.16 \pm 0.19$  \\
\hline
$\gamma_{PN}$                                              & $> 0.999943$ (95\% CL)            & $> 0.999981$ (95\% CL)            & $> 0.999968$ (95\% CL)  \\
$\beta_{PN}$                                               & $< 1.0000048$ (95\% CL)           & $< 1.0000015$ (95\% CL)           & $< 1.0000027$ (95\% CL)  \\
$\delta G_\mathrm{N}/G_\mathrm{N}$ (z=0)                   & $> -0.052$ (95\% CL)              & $> -0.018$ (95\% CL)              & $> -0.030$ (95\% CL)  \\
$10^{13} \dot{G}_\mathrm{N}/G_{\rm N}$ (z=0) [yr$^{-1}$]   & $> -7.5\times 10^{-9}$ (95\% CL)  & $> -2.5\times 10^{-9}$ (95\% CL)  & $> -4.3\times 10^{-9}$ (95\% CL)  \\
$G_\mathrm{N}/G$ (z=0)                                     & $> 0.999975$ (95\% CL)            & $> 0.999991$ (95\% CL)            & $> 0.999984$ (95\% CL)  \\
\hline
$\Omega_{\rm m}$                        & $0.299^{+0.014}_{-0.011}$  & $0.3070 \pm 0.0066$  & $0.2929 \pm 0.0062$  \\
$\sigma_8$                              & $0.827^{+0.011}_{-0.013}$  & $0.8204 \pm 0.0099$  & $0.8391 \pm 0.0095$  \\
$r_s$ [Mpc]                             & $149.5 \pm 2.0$            & $149.3 \pm 2.0$      & $145.5 \pm 1.6$  \\
\hline
$\Delta \chi^2$                         & $1.4$ & $-0.2$ & $-3.8$  \\
\hline
\hline
\end{tabular}}
\caption{\label{tab:cc_Neff} 
Constraints on main and derived parameters (at 68\% CL if not otherwise stated) considering 
P18 in combination with BAO and BAO + R19 for the CC+$N_{\rm eff}$ model.}
\end{table*}

    \newpage

\subsection{Degeneracy with the neutrino sector: $m_\nu$}

\begin{table*}[h!]
{\small
\centering
\begin{tabular}{l|ccc}
\hline
\hline
                                         & P18 & P18 + BAO & P18 + BAO + R19  \\
\hline
$\omega_{\rm b}$                         & $0.02239 \pm 0.00017$         & $0.02241 \pm 0.00014$                 & $0.02247 \pm 0.00013$  \\
$\omega_{\rm c}$                         & $0.1205 \pm 0.0013$           & $0.1203 \pm 0.0011$                   & $0.1203 \pm 0.0012$  \\
$H_0$ [km s$^{-1}$Mpc$^{-1}$]            & $68.5 \pm 1.8\ (2.4\sigma)$   & $68.66_{-0.87}^{+0.69}\ (3.4\sigma)$  & $70.12 \pm 0.81\ (2.4\sigma)$  \\
$\tau$                                   & $0.0567^{+0.0065}_{-0.0082}$  & $0.0564_{-0.0080}^{+0.0066}$          & $0.0572^{+0.0063}_{-0.0080}$  \\
$\ln \left(  10^{10} A_{\rm s} \right)$  & $3.052_{+0.013}^{-0.016}$     & $3.051_{+0.013}^{-0.016}$             & $3.054^{+0.013}_{-0.016}$  \\
$n_{\rm s}$                              & $0.9668 \pm 0.0053$           & $0.9672 \pm 0.0038$                   & $0.9700 \pm 0.0038$  \\
$\zeta_{\rm IG}$                         & $< 0.0037$ (95\% CL)          & $< 0.0030$ (95\% CL)                  & $0.0026^{+0.0010}_{-0.0013}$  \\
$m_\nu$ [eV]                             & $< 0.31$ (95\% CL)            & $< 0.17$ (95\% CL)                    & $< 0.19$ (95\% CL)  \\
\hline
$\xi$                                                     & $< 0.00094$ (95\% CL)  & $< 0.00076$ (95\% CL)  & $0.00065 \pm 0.00057$ (95\% CL)  \\
$\gamma_{PN}$                                             & $> 0.9963$ (95\% CL)   & $> 0.9970$ (95\% CL)   & $0.9974^{+0.0013}_{-0.0010}$  \\
$\delta G_\mathrm{N}/G_\mathrm{N}$ (z=0)                  & $> -0.027$ (95\% CL)   & $> -0.022$ (95\% CL)   & $-0.0190_{-0.0075}^{+0.0093}$  \\
$10^{13} \dot{G}_\mathrm{N}/G_{\rm N}$ (z=0) [yr$^{-1}$]  & $> -1.1$ (95\% CL)     & $> -0.93$ (95\% CL)    & $-0.78^{+0.39}_{-0.31}$  \\
$G_\mathrm{N}/G$ (z=0)                                    & $> 0.9981$ (95\% CL)   & $> 0.9985$ (95\% CL)   & $0.9987^{+0.00064}_{-0.00051}$  \\
\hline
$\Omega_{\rm m}$                        & $0.306_{-0.018}^{+0.015}$  & $0.3029 \pm 0.0076$        & $0.2905 \pm 0.0068$  \\
$\sigma_8$                              & $0.815_{-0.014}^{+0.025}$  & $0.821_{-0.010}^{+0.014}$  & $0.832 \pm 0.013$  \\
$r_s$ [Mpc]                             & $146.18_{-0.38}^{+0.78}$   & $146.31_{-0.37}^{+0.71}$   & $145.56_{-0.69}^{+0.78}$  \\
\hline
$\Delta \chi^2$                         & $3.0$ & $0.2$ & $-3.3$  \\
\hline
\hline
\end{tabular}}
\caption{Constraints on main and derived parameters (at 68\% CL if not otherwise stated) considering 
P18 in combination with BAO and BAO + R19 for the IG+$m_\nu$ model.}
\label{tab:ig_mnu} 
\end{table*}

\begin{table*}[h!]
{\small
\centering
\begin{tabular}{l|ccc}
\hline
\hline
                                         & P18 & P18 + BAO & P18 + BAO + R19  \\
\hline
$\omega_{\rm b}$                         & $0.02240 \pm 0.00016$         & $0.02242 \pm 0.00013$                 & $0.02252 \pm 0.00013$  \\
$\omega_{\rm c}$                         & $0.1203 \pm 0.0013$           & $0.12011 \pm 0.00097$                 & $0.1197 \pm 0.0010$  \\
$H_0$ [km s$^{-1}$Mpc$^{-1}$]            & $68.0 \pm 1.4\ (3.0\sigma)$   & $68.31_{-0.69}^{+0.62}\ (3.7\sigma)$  & $69.62 \ 0.71\ (2.8\sigma)$  \\
$\tau$                                   & $0.0563_{-0.0080}^{+0.0063}$  & $0.0564_{-0.0077}^{+0.0065}$          & $0.0576^{+0.0067}_{-0.0077}$  \\
$\ln \left(  10^{10} A_{\rm s} \right)$  & $3.051_{-0.016}^{+0.013}$     & $3.047_{-0.015}^{+0.013}$             & $3.054^{+0.013}_{-0.016}$  \\
$n_{\rm s}$                              & $0.9674 \pm 0.0053$           & $0.9681 \pm 0.0043$                   & $0.9720 \pm 0.0041$  \\
$N_{\rm pl}$ [M$_{\rm pl}$]              & $< 1.000026$ (95\% CL)        & $< 1.000024$ (95\% CL)                & $1.000019^{+0.000017}_{-0.000018}$ (95\% CL)  \\
$m_\nu$ [eV]                             & $< 0.28$ (95\% CL)            & $< 0.16$  (95\% CL)                   & $< 0.13$ (95\% CL)  \\
\hline
$\gamma_{PN}$                                             & $> 0.999926$ (95\% CL)             & $> 0.999924$ (95\% CL)            & $0.9999192_{-0.000011}^{+0.000009}$ (95\% CL)  \\
$\beta_{PN}$                                              & $< 1.0000021$ (95\% CL)            & $< 1.0000020$ (95\% CL)           & $< 1.0000030$ (95\% CL)  \\
$\delta G_\mathrm{N}/G_\mathrm{N}$                        & $> -0.024$ (95\% CL)               & $> -0.023$ (95\% CL)              & $-0.0181^{+0.0099}_{-0.0082}$  \\
$10^{13} \dot{G}_\mathrm{N}/G_{\rm N}$ (z=0) [yr$^{-1}$]  & $> -3.6\times 10^{-9}$ (95\% CL)   & $> -3.3\times 10^{-9}$ (95\% CL)  & $(-2.7^{+1.5}_{-1.2})\times 10^{-9}$  \\
$G_\mathrm{N}/G$ (z=0)                                    & $> 0.999987$ (95\% CL)             & $> 0.9999988$ (95\% CL)           & $0.9999904^{+0.0000054}_{-0.0000043}$  \\
\hline
$\Omega_{\rm m}$                        & $0.309^{+0.011}_{-0.015}$ & $0.3047 \pm 0.0067$ & $0.2935 \pm 0.0064$  \\
$\sigma_8$                              & $0.814 \pm 0.010$ & $0.820_{-0.010}^{+0.013}$   & $0.831 \pm 0.012$  \\
$r_s$ [Mpc]                             & $146.52_{-0.34}^{+0.47}$     & $146.58^{+0.47}_{-0.31}$     & $146.26^{+0.55}_{-0.48}$  \\
\hline
$\Delta \chi^2$                         & $3.0$ & $0.0$ &  $-1.5$ \\
\hline
\hline
\end{tabular}}
\caption{\label{tab:cc_mnu} 
Constraints on main and derived parameters (at 68\% CL if not otherwise stated) considering 
P18 in combination with BAO and BAO + R19 for the CC+$m_\nu$ model.}
\end{table*}

    \newpage

\subsection{Degeneracy with the neutrino sector: ($N_{\rm eff},\ m_\nu$)}

\begin{table*}[h!]
{\small
\centering
\begin{tabular}{l|ccc}
\hline
\hline
                                          & P18  &  P18 + BAO  &  P18 + BAO + R19  \\
\hline
$\omega_{\rm b}$                          & $0.02218 \pm 0.00022$              & $0.02220^{+0.00022}_{-0.00019}$  & $0.02250 \pm 0.00020$  \\
$\omega_{\rm c}$                          & $0.1162 \pm 0.0034$                & $0.1164 \pm 0.0031$              & $0.1208 \pm 0.0030$  \\
$H_0$ [km s$^{-1}$Mpc$^{-1}$]             & $67.7^{+2.0}_{-2.4}\ (2.6\sigma)$  & $67.6 \pm 1.2 (3.5\sigma)$       & $70.25 \pm 0.92\ (2.2\sigma)$  \\
$\tau$                                    & $0.0556^{+0.0065}_{-0.0083}$       & $0.0554_{-0.0073}^{+0.0065}$     & $0.0576^{+0.0063}_{-0.0081}$  \\
$\ln \left(  10^{10} A_{\rm s} \right)$   & $3.039_{+0.016}^{-0.018}$          & $3.039 \pm 0.016$                & $3.056 \pm 0.016$  \\
$n_{\rm s}$                               & $0.9577 \pm 0.0086$                & $0.9582 \pm 0.0076$              & $0.9710 \pm 0.0071$  \\
$\zeta_{\rm IG}$                          & $< 0.0070$ (95\% CL)               & $< 0.0047$ (95\% CL)             & $< 0.0053$ (95\% CL)  \\
$m_\nu$ [eV]                              & $< 0.26$ (95\% CL)                 & $< 0.19$   (95\% CL)             & $< 0.19$ (95\% CL)  \\
$N_{\rm eff}$                             & $2.74 \pm 0.22$                    & $2.77 \pm 0.20$                  & $3.08 \pm 0.20$  \\
\hline
$\xi$                                                     & $< 0.0018$ (95\% CL)  & $< 0.0012$ (95\% CL)  & $< 0.0013$ (95\% CL)  \\
$\gamma_{PN}$                                             & $> 0.9931$ (95\% CL)  & $> 0.9954$ (95\% CL)  & $> 0.9948$ (95\% CL)  \\
$\delta G_\mathrm{N}/G_\mathrm{N}$ (z=0)                  & $> -0.050$ (95\% CL)  & $> -0.034$ (95\% CL)  & $> -0.038$  (95\% CL)  \\
$10^{13} \dot{G}_\mathrm{N}/G_{\rm N}$ (z=0) [yr$^{-1}$]  & $> -2.0$ (95\% CL)    & $> -1.4$ (95\% CL)    & $> 1.6$ (95\% CL)  \\
$G_\mathrm{N}/G$ (z=0)                                    & $> 0.9966$ (95\% CL)  & $> 0.9977$ (95\% CL)  & $> 0.9974$ (95\% CL)  \\
\hline
$\Omega_{\rm m}$                        & $0.303_{-0.019}^{+0.022}$  & $0.3035 \pm 0.0081$        & $0.2904 \pm 0.0069$  \\
$\sigma_8$                              & $0.814^{+0.025}_{-0.019}$  & $0.815_{-0.012}^{+0.015}$  & $0.833^{+0.013}_{-0.011}$  \\
$r_s$ [Mpc]                             & $148.6 \pm 1.9$            & $148.6 \pm 1.8$            & $145.3 \pm 1.6$  \\
\hline
$\Delta \chi^2$                         & $1.1$ & $0.5$ & $-2.5$  \\
\hline
\hline
\end{tabular}}
\caption{Constraints on main and derived parameters (at 68\% CL if not otherwise stated) considering 
P18 in combination with BAO and BAO + R19 for the IG+$N_{\rm eff}$+$m_\nu$ model.}
\label{tab:ig_Neff_mnu} 
\end{table*}

\begin{table*}[h!]
{\small
\centering
\begin{tabular}{l|ccc}
\hline
\hline
                                         & P18 & P18 + BAO & P18 + BAO + R19  \\
\hline
$\omega_{\rm b}$                         & $0.02217 \pm 0.00022$          & $0.02222 \pm 0.00020$         & $0.02257 \pm 0.00018$  \\
$\omega_{\rm c}$                         & $0.1158 \pm 0.0034$            & $0.1158 \pm 0.0032$           & $0.1212 \pm 0.0031$  \\
$H_0$ [km s$^{-1}$Mpc$^{-1}$]            & $66.7 \pm 1.8\ (3.2\sigma)$    & $67.2 \pm 1.1 (3.8\sigma)$    & $69.96 \pm 0.93\ (2.1\sigma)$  \\
$\tau$                                   & $0.0554^{+0.0064}_{-0.0076}$   & $0.0556_{-0.0075}^{+0.0063}$  & $0.0577^{+0.0069}_{-0.0082}$  \\
$\ln \left(  10^{10} A_{\rm s} \right)$  & $3.039 \pm 0.017$              & $3.039 \pm 0.016$             & $3.057 \pm 0.016$   \\
$n_{\rm s}$                              & $0.9582 \pm 0.0084$            & $0.9596 \pm 0.0074$           & $0.9745 \pm 0.0064$  \\
$N_{pl}$ [M$_{\rm pl}$]                  & $< 1.000050$ (95\% CL)         & $< 1.000042$ (95\% CL)        & $< 1.000040$ (95\% CL)  \\
$m_\nu$ [eV]                             & $< 0.26$ (95\% CL)             & $< 0.17$   (95\% CL)          & $< 0.14$ (95\% CL)  \\
$N_{\rm eff}$                            & $2.73 \pm 0.21$                & $2.75 \pm 0.21$               & $3.14 \pm 0.20$  \\
\hline
$\gamma_{PN}$                                             & $> 0.999950$ (95\% CL)            & $> 0.9958$ (95\% CL)              & $> 0.9960$ (95\% CL)  \\
$\beta_{PN}$                                              & $< 1.0000041$ (95\% CL)           & $< 1.0000035$ (95\% CL)           & $< 1.0000033$ (95\% CL)  \\
$\delta G_\mathrm{N}/G_\mathrm{N}$ (z=0)                  & $> -0.046$ (95\% CL)              & $> -0.040$ (95\% CL)              & $> -0.037$  (95\% CL)  \\
$10^{13} \dot{G}_\mathrm{N}/G_{\rm N}$ (z=0) [yr$^{-1}$]  & $> -6.7\times 10^{-9}$ (95\% CL)  & $> -5.7\times 10^{-9}$ (95\% CL)  & $> -5.5\times 10^{-9}$ (95\% CL)  \\
$G_\mathrm{N}/G$ (z=0)                                    & $> 0.999975$ (95\% CL)            & $> 0.999979$ (95\% CL)            & $> 0.999980$ (95\% CL)  \\
\hline
$\Omega_{\rm m}$                        & $0.310_{-0.018}^{+0.016}$  & $0.3056 \pm 0.0074$        & $0.2939 \pm 0.0064$  \\
$\sigma_8$                              & $0.808^{+0.024}_{-0.015}$  & $0.814_{-0.011}^{+0.015}$  & $0.833 \pm 0.012$  \\
$r_s$ [Mpc]                             & $149.3^{+1.8}_{-2.1}$      & $149.2 \pm 1.9$            & $145.4 \pm 1.7$  \\
\hline
$\Delta \chi^2$                         & $3.0$ & $0.4$ & $-0.6$  \\
\hline
\hline
\end{tabular}}
\caption{Constraints on main and derived parameters (at 68\% CL if not otherwise stated) considering 
P18 in combination with BAO and BAO + R19 for the CC+$N_{\rm eff}$+$m_\nu$ model.}
\label{tab:cc_Neff_mnu} 
\end{table*}

    \newpage



\end{document}